\documentclass[fleqn,usenatbib,useAMS]{mnras}

\usepackage{newtxtext,newtxmath}

\usepackage[T1]{fontenc}

\DeclareRobustCommand{\VAN}[3]{#2}
\let\VANthebibliography\thebibliography
\def\thebibliography{\DeclareRobustCommand{\VAN}[3]{##3}\VANthebibliography}
\usepackage{ae,aecompl}

\usepackage{graphicx}	
\usepackage{amsmath}	
\usepackage{xcolor}
\usepackage{multicol}
\usepackage{xspace}
\usepackage{booktabs} 
\usepackage{soul}
\usepackage{eso-pic}

\newcommand{\maglim}{\textsc{MagLim}\xspace}
\newcommand{\dnf}{\textsc{DNF}\xspace}
\newcommand{\tjpcov}{\textsc{TJPCov}\xspace}
\newcommand{\namaster}{\textsc{NaMASTER}\xspace}
\newcommand{\cosmolike}{\textsc{CosmoLike}\xspace}
\newcommand{\redmagic}{\textsc{redMaGiC}\xspace}
\newcommand{\metacalibration}{\textsc{METACALIBRATION}\xspace}
\newcommand{\healpix}{\textsc{HEALPix}\xspace}
\newcommand{\halofit}{\textsc{HALOFIT}\xspace}
\newcommand{\flask}{\textsc{FLASK}\xspace}
\newcommand{\cosmosis}{\textsc{CosmoSiS}\xspace}
\newcommand{\LCDM}{$\Lambda$CDM\xspace}
\newcommand{\wCDM}{$w$CDM\xspace}
\newcommand{\gold}{\textsc{Gold}\xspace}

\definecolor{Forestgreen}{rgb}{0.133, 0.545, 0.133}

\AddToShipoutPictureBG*{%
  \AtPageUpperLeft{%
    \hspace{0.75\paperwidth}%
    \raisebox{-3.5\baselineskip}{%
      \makebox[0pt][l]{\textnormal{DES-2024-0825}}
}}}%

\AddToShipoutPictureBG*{%
  \AtPageUpperLeft{%
    \hspace{0.75\paperwidth}%
    \raisebox{-4.5\baselineskip}{%
      \makebox[0pt][l]{\textnormal{FERMILAB-PUB-24-0289-PPD}}
}}}%

\title[DES Y3 Harmonic Space 2$\times$2pt]{Dark Energy Survey Year 3 Results: Cosmology from galaxy clustering and galaxy-galaxy lensing in harmonic space}

\author[DES Collaboration]{
\parbox{\textwidth}{
\Large
L.~Faga,$^{1,2}$
F.~Andrade-Oliveira,$^{3}$
H.~Camacho,$^{4,5,2}$
R.~Rosenfeld,$^{6,2}$
M.~Lima,$^{1,2}$
C.~Doux,$^{7,8}$
X.~Fang,$^{9,10}$
J.~Prat,$^{11,12}$
A.~Porredon,$^{13}$
M.~Aguena,$^{2}$
% ...
\begin{center} (DES Collaboration) \end{center}
}
\vspace{-1cm}
\\
}

\date{Accepted XXX. Received YYY; in original form ZZZ}

\pubyear{2015}

\begin{document}
\label{firstpage}
\pagerange{\pageref{firstpage}--\pageref{lastpage}}
\maketitle

\begin{abstract}
	
We present the joint tomographic analysis of galaxy-galaxy lensing and galaxy clustering in harmonic space, using galaxy catalogues from the first three years of observations by the Dark Energy Survey (DES Y3). We utilise the \redmagic and \maglim catalogues as lens galaxies and the \metacalibration catalogue as source galaxies. The measurements of angular power spectra are performed using the pseudo-$C_\ell$ method, and our theoretical modelling follows the fiducial analyses performed by DES Y3 in configuration space, accounting for galaxy bias, intrinsic alignments, magnification bias, shear magnification bias and photometric redshift uncertainties. We explore different approaches for scale cuts based on non-linear galaxy bias and baryonic effects contamination. Our fiducial covariance matrix is computed analytically, accounting for mask geometry in the Gaussian term, and including non-Gaussian contributions and super-sample covariance terms. To validate our harmonic space pipelines and covariance matrix, we used a suite of 1800 log-normal simulations. We also perform a series of stress tests to gauge the robustness of our harmonic space analysis.
In the $\Lambda$CDM model, the clustering amplitude $S_8 =\sigma_8(\Omega_m/0.3)^{0.5}$ is constrained to $S_8 = 0.704\pm 0.029$ and $S_8 = 0.753\pm 0.024$ (68\% C.L.) for the \redmagic and \maglim catalogues, respectively.
For the \wCDM, the dark energy equation of state is constrained to $w = -1.28 \pm 0.29$ and $w = -1.26^{+0.34}_{-0.27}$,
for \redmagic and \maglim catalogues, respectively.
These results are compatible with the corresponding DES Y3 results in configuration space and pave the way for harmonic space analyses using the DES Y6 data.
\end{abstract}




\section{Introduction}
\label{sec:intro}

The accelerating expansion of the Universe, first discovered through the analysis of supernova light curves \citep{Perlmutter1997,Perlmutter1999,Riess1998}, has received significant further evidence from a variety of complementary observations.
These include measurements of properties of the cosmic microwave background (CMB) \citep{WMAP2013,Planck2020,ACT2024} and the large scale structure (LSS) of the universe (see \citealp{ReviewLSSProbes2013} for a recent review).
Analysis from these different observations have statistically converged during the recent decades upon a concordance cosmological model, the $\Lambda$CDM model.
This model describes a spatially flat Universe composed of roughly 30\% of visible and cold dark matter (CDM), and 70\% dark energy.
This dark energy is responsible for the accelerated expansion and is consistent with a cosmological constant ($\Lambda$), but its physical nature remains as an open problem.

Recent progress has been significant in enhancing the data quality and quantity, along with improved techniques, for extracting cosmological information from various LSS probes of cosmic acceleration.
Notably, the distribution of matter traced by galaxy positions (galaxy clustering) and the weak gravitational lensing distortion it induces on the shapes of distant galaxies (cosmic shear) have proven to be powerful tools for constraining cosmological models \citep{desy1_3x2pt,Abbott_2022,KiDS2021,eBOSS2021,Dalal2023,Li2023}. 
These techniques are considered to be amongst the key scientific drivers of current Stage-III observational programmes and will continue to be for the next generation of experiments \citep{LSST2019,EUCLID2011}.
The Dark Energy Survey [DES\footnote{\url{http://www.darkenergysurvey.org/}} \citet{2005astro.ph.10346T}]
is the largest stage-III imaging survey today, covering almost 5000 deg$^2$ of the southern sky. 
One of the key cosmological results obtained from the large scale structure and weak lensing data from the first three years of observations (DES Y3) consisted of the analysis of three two-point angular correlation functions (3$\times$2pt) arising from the clustering and gravitational shear of galaxies \citep{Abbott_2022}.
The final analyses using the full 6 years of data (DES Y6) are well under way.

In addition to these main results obtained from the configuration/real space angular correlation functions, complementary analyses using their Fourier/Harmonic counterpart, the two-point angular power spectra, were also conducted within DES.
The angular power spectra of the data collected in the first year (DES Y1) were studied for galaxy clustering \citep{Andrade_Oliveira_2021} and cosmic shear \citep{Camacho_2021} whereas \cite{Doux_2022} obtained cosmological constraints from the analysis of cosmic shear in harmonic space for DES Y3.
These results add up to the ongoing generation of observations from Stage-III dark energy experiments, having as a main scientific goal a better understanding of the nature of cosmic acceleration.
In the context of combined multiprobe analysis from weak gravitational lensing and galaxy clustering,
the Kilo-Degree Survey (KiDS\footnote{\url{http://kids.strw.leidenuniv.nl/}}) has released results combining its cosmic shear observations with galaxy clustering from the overlapping Baryon Oscillation Spectroscopic Survey (BOSS\footnote{\url{https://www.sdss.org/surveys/boss/}}) and the 2-degree Field Lensing Survey (2dFLenS\footnote{\url{https://2dflens.swin.edu.au/}}) in configuration space
\citep{Joachimi_KiDSxBOSS2dFLenS-Methods,Heymans_KiDSxBOSS2dFLenS}.
Analogously, the Hyper Suprime-Cam (HSC\footnote{\url{https://www.naoj.org/Projects/HSC/}}) has also cross-correlated its weak lensing signal measurements with clustering from BOSS \citep{More_HSCxBOSS}, but based in configuration space.
In a complementary way, the cosmic shear angular power spectrum was also recently measured from the three-year galaxy shear catalogue from the HSC imaging survey \citep{Dalal2023,Li2023}.
A reanalysis of HSC data only, which combines the galaxy clustering and weak lensing signals in harmonic space, is currently in development \citep{SanchezCid_2024}.

In this paper we present, for the first time within the DES, results from a combination of two angular power spectra in harmonic space, namely the two-point correlation of galaxy positions, $\langle \delta_g \delta_g \rangle$, and the cross-correlation of galaxy positions and shapes, $\langle \delta_g \gamma \rangle$, which we refer to as 2$\times$2pt.
The proposed methodology is tested for internal consistency and robustness and the results are compared to the ones obtained from the configuration space analyses on the same dataset \citep{2x2_redmagic,2x2_maglim}.
For the \maglim sample, we present results that utilise the first four tomographic bins (employed in the fiducial results of \citealp{2x2_maglim}), and with all six tomographic bins, and discuss the consistency of the results across these different redshift ranges.
This work represents an important milestone probing the robustness of the different analyses of DES data, being a completely independent data reduction to a different set of summary statistics. The presented methodology has its own estimators, covariances and modelling, constraining the survey information in a unique way. This paves the way for a full 3$\times$2pt analysis in harmonic space using the final DES Y6 data set, as well as for future analyses of next-generation cosmological surveys.

This paper is organised as follows. In Section~\ref{sec:data} we review the data used, namely the two catalogues for the lens galaxies, \redmagic \citep{2x2_redmagic} and \maglim \citep{2021PhRvD.103d3503P}, and the \metacalibration catalogue for the source galaxies \citep*{y3-shapecatalog}.
In this section, we also describe the generation of log-normal simulations used to validate the estimated covariance matrix and assess the accuracy of our cosmological analysis pipeline.
Section~\ref{sec:methodology} describes the theoretical modelling of the angular power spectra, accounting for galaxy bias, intrinsic alignments, magnification bias, shear magnification bias and photometric redshift uncertainties.
Scale cuts are devised to mitigate small-scale uncertainties and a likelihood analysis is presented and tested against simulated data vectors.
Our results for $\Lambda$CDM and $w$CDM models are presented in Section~\ref{sec:results} along with internal consistency and robustness tests, and our conclusions are summarised in Section~\ref{sec:conc}.


\begin{figure*}
\centering
\includegraphics[width=0.33\textwidth]{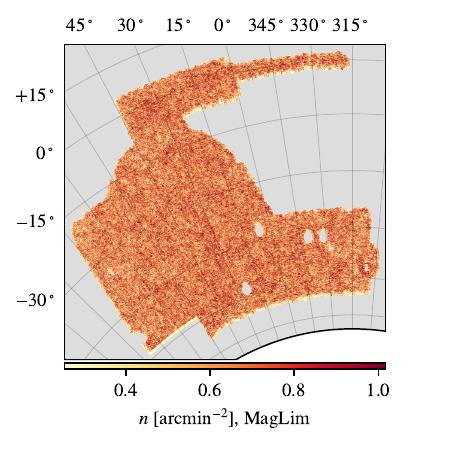}
\includegraphics[width=0.33\textwidth]{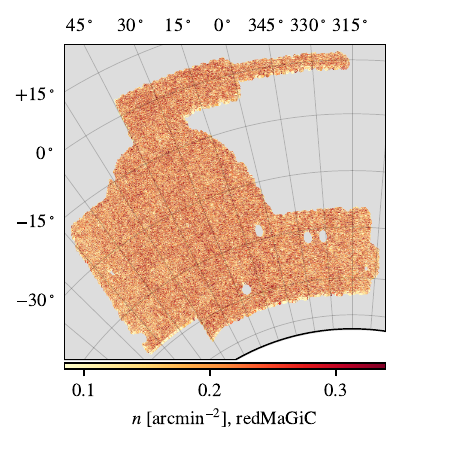}
\includegraphics[width=0.33\textwidth]{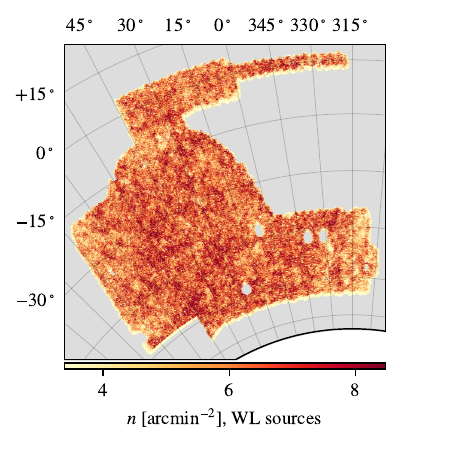} \\
\includegraphics[width=0.33\textwidth]{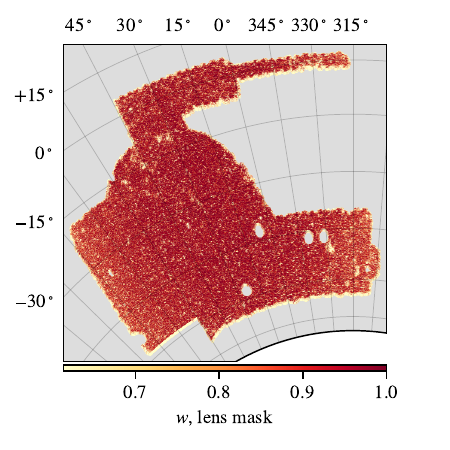}
\includegraphics[width=0.33\textwidth]{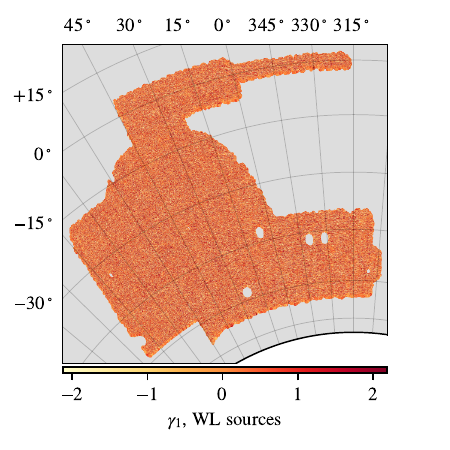}
\includegraphics[width=0.33\textwidth]{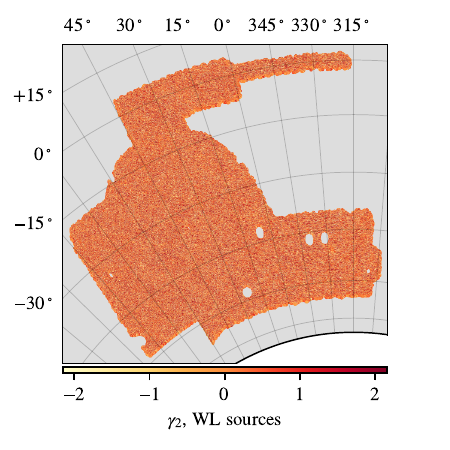}
\caption{
    The maps of the different DES Y3 catalogues used in this work. 
    Top: the number densities of the full \redmagic, \maglim and \metacalibration samples, without tomographic selection.
    Bottom left: The covered fraction of the sky common to all samples, used as the mask for the lens samples.
    Bottom middle and right: the two components of shear ellipticities in the \metacalibration sample.
}
\label{fig:catalogues}
\end{figure*}

\section{Data}
\label{sec:data}

The DES is a photometric galaxy survey that imaged about one-fourth of the southern sky in five optical filters: $g, r, i, z$, and $Y$, collecting data from more than 500 million galaxies.
Using the Dark Energy Camera (DECam) on the Blanco telescope at the Cerro Tololo Inter-American Observatory (CTIO) in Chile, DES completed its observations in January 2019 after 6 years of operations.
In this paper, we utilise data from the initial three-year observation period (Y3), spanning August 2013 to February 2016.

\subsection{DES Y3 catalogues}

The DES Y3 photometric data set resulted in a catalogue encompassing nearly 390 million objects over an area of 4946 deg$^2$ of the sky, with a depth reaching a signal-to-noise ratio of approximately 10 for extended objects up to an AB system magnitude in the $i$ band of approximately 23.
Its detailed selection is referred to as the \gold catalogue, and is described in detail in \cite{y3-gold}.
To mitigate the influence of astrophysical foregrounds and known data processing artefacts on subsequent cosmological analyses, additional masking was implemented.
This process resulted in a final catalogue encompassing a reduced area of 4143 deg$^2$.
This dataset was used to further select two distinct lens samples and one source sample of galaxies.
The source sample serves as the background population for weak lensing analysis, while the lens samples are employed for both weak lensing, as foreground population, and galaxy clustering analyses.

\begin{figure}
  \centering
  \includegraphics[width=\columnwidth]{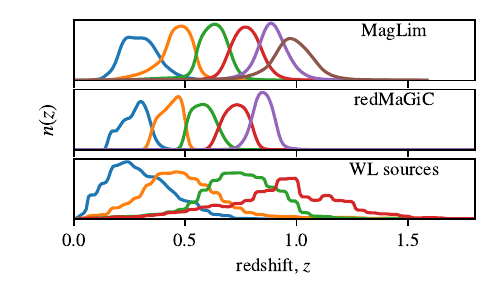}
  \caption{Normalised redshift distributions for the DES Y3 lens and source catalogues. 
  Each panel corresponds to a sample and each curve corresponds to a tomographic bin selection.
  }
  \label{fig:nz}
\end{figure}

The first lens sample used in the DES Y3 analyses was constructed using the \redmagic algorithm (outlined in \citealp{sv-redmagic}).
This algorithm is specifically designed to identify luminous red galaxies (LRGs) with minimal uncertainties in their photometric redshifts.
It achieves this by leveraging the well-established magnitude-colour-redshift relationship exhibited by red-sequence galaxies.
The resulting selection on the DES Y3 data, the so-called \redmagic sample, consists of 2.6 million galaxies, which are divided into five tomographic bins described in Table \ref{tab:LensSamples}.
The full \redmagic catalogue's angular density is depicted in the upper-central panel of Figure \ref{fig:catalogues}.
The redshift distribution of each bin is shown in Figure \ref{fig:nz}.
For more details about the \redmagic algorithm, the \redmagic sample, and its comparison with other samples we refer the reader to \cite*{sv-redmagic} and \cite{2x2_redmagic}, respectively.

The second, and fiducial, lens sample used for the DES Y3 analyses, the so-called \maglim sample, was optimised in \cite{2021PhRvD.103d3503P} to maximise its constraining power for combined galaxy clustering and galaxy-galaxy lensing.
\maglim is a magnitude-limited sample defined by a magnitude cut in the $i$-band that is linearly dependent on the photometric redshift, which allows including more galaxies at higher redshift.
The photometric redshift estimation for the \maglim sample used the Directional Neighborhood Fitting (\dnf) algorithm~\citep{DNF}.
In order to remove stellar contamination from binary stars and other bright objects, the \maglim selection applies a further lower magnitude cut of $i > 17.5$.
The resultant sample, comprising 10.7 million galaxies, is divided into the six tomographic bins detailed in Table \ref{tab:LensSamples}\footnote{We note bin edges in Table \ref{tab:LensSamples} are slightly modified from \cite{2021PhRvD.103d3503P} and shown in Figure \ref{fig:nz} to improve the photometric redshift calibration (see \cite{2x2_maglim} for further discussion).}
and comprises a wider redshift distribution and 3.5 times more galaxies compared to the \redmagic sample, resulting in tighter cosmological parameter constraints \citep{2x2_maglim}.
The angular density of the full \maglim catalogue is shown on the top-left panel of Figure \ref{fig:catalogues}.

Statistically significant correlations were found between the galaxy number density of both the \redmagic and \maglim lens samples and various observational survey properties.
Such correlation imprints a non-cosmological bias into the galaxy clustering signal for the lens samples.
We account for those biases by applying weights to each galaxy corresponding to the inverse of the estimated angular selection function.
The computation and validation of these weights, for both lens samples, is described in \cite{2022MNRAS.511.2665R}.

Finally, we used as source sample the DES Y3 weak lensing shape catalogue \citep*{y3-shapecatalog}.
The source shapes are measured using the \metacalibration method \citep{2014MNRAS.444L..25S,2017ApJ...841...24S,2017arXiv170202600H}. 
This method offers a self-calibrating approach to correct for biases in the galaxy shear estimation.
This is achieved by applying an iterative process where a single elliptical Gaussian model is fitted to artificially sheared replicas of each galaxy.
This procedure results in the construction of a shear response matrix, denoted as $R = R_\gamma + R_S$, which encapsulates two distinct components: $R_\gamma$ representing the response of the shape estimator and $R_S$ representing the response of object selection.
The calibrated shear measurements are then obtained by multiplying the estimated galaxy ellipticities by the inverse of this matrix \citep*{y3-shapecatalog}.

For the DES Y3 shape catalogue, the \metacalibration method improves the accuracy of galaxy shape measurements over its DES Y1 counterpart \citep{2018MNRAS.481.1149Z} by incorporating per galaxy inverse variance weights based on signal-to-noise ratio and size \citep*{y3-shapecatalog}, better astrometric methods \citep{y3-gold}, and better point-spread function (PSF) estimation \citep{2021MNRAS.501.1282J}. 
On top of that, shear biases sourced by object blending, not taken into account by \metacalibration, were calibrated using image simulations in \cite{2022MNRAS.509.3371M}.
The resultant shape catalogue comprises 100.2 million galaxies covering the same area as the lens catalogues.
This translates to an effective angular galaxy density of $5.59$ arcmin$^{-2}$, as defined by \cite{neff-heymans}. Furthermore, the catalogue exhibits an effective shape noise $\sigma_e = 0.261$ per ellipticity component.
The catalogue is further subdivided into four tomographic bins, selecting in photometric redshift estimates that rely on the Self-Organizing Map Photometric Redshift (SOMPZ) method as described by \citet*{y3-sompz}, each possessing a normalised redshift distribution as illustrated in Figure \ref{fig:nz} and properties detailed in Table \ref{tab:SourceSample}.

\begin{table}
    \caption{
    The DES Y3 lens samples specifications used in this work.
    This table shows the photometric redshift selection, total number of galaxies selected, ($N_{\rm gal}$), effective angular number density of galaxies in ${\rm arcmin}^{-2}$, ($n_{\rm eff}$), and magnification bias, ($C_{\rm g}$),
    as measured in \protect\cite*{y3-2x2ptmagnification}.
    The survey property systematic weights have been accounted for in the effective angular number density following Equation \eqref{eq:WeightedAngularNumberDensity}.
        } 
    \centering
    \begin{tabular}{cccc}
        \hline\hline
            redshift bin & $N_{\mathrm{gal}}$ & $n_\mathrm{eff}$ & $C_{\rm g}$ \\
        \hline
        \multicolumn{4}{c}{\redmagic lens sample} \\
        \hline
        $0.15 < z_\mathrm{ph} < 0.35$ & 330 243 & 0.022 & 1.3134  \\
        $0.35 < z_\mathrm{ph} < 0.50$ & 571 551 & 0.038 & -0.5179 \\
        $0.50 < z_\mathrm{ph} < 0.65$ & 872 611 & 0.058 & 0.3372  \\
        $0.65 < z_\mathrm{ph} < 0.80$ & 442 302 & 0.029 & 2.2515  \\
        $0.80 < z_\mathrm{ph} < 0.90$ & 377 329 & 0.025 & 1.9667  \\
        \hline
        \multicolumn{4}{c}{\maglim lens sample} \\
        \hline
        $0.20 < z_\mathrm{ph} < 0.40$ & 2 236 473 &  0.150 & 1.2143 \\
        $0.40 < z_\mathrm{ph} < 0.55$ & 1 599 500 &  0.107 & 1.1486 \\
        $0.55 < z_\mathrm{ph} < 0.70$ & 1 627 413 &  0.109 & 1.8759 \\
        $0.70 < z_\mathrm{ph} < 0.85$ & 2 175 184 &  0.146 & 1.9694 \\
        $0.85 < z_\mathrm{ph} < 0.95$ & 1 583 686 &  0.106 & 1.7805 \\
        $0.95 < z_\mathrm{ph} < 1.05$ & 1 494 250 &  0.100 & 2.4789 \\
        \hline\hline
    \end{tabular}
    \label{tab:LensSamples}
\end{table}

\begin{table}
    \caption{
    The DES Y3 source sample properties.
    Similar to Table \ref{tab:LensSamples}, this table shows the photometric redshift selection,
    the effective galaxy number density, ($n_{\rm eff}$) in arcmin$^{-2}$,
    the shape-noise per-component, ($\sigma_e$) and
    the mean shear response, ($\langle R_\gamma \rangle$), and the mean selection response, ($\langle R_S \rangle$).
    }  
    \centering
    \begin{tabular}{ccccc}
        \hline\hline
	redshift bin & $n_\mathrm{eff}$ & $\sigma_e$ & $\langle R_\gamma \rangle$  & $\langle R_S \rangle$ \\
        \hline 
        $0.0 < z_\mathrm{ph} < 0.36$  & 1.476 & 0.243  & 0.7636 & 0.0046 \\
        $0.36 < z_\mathrm{ph} < 0.63$ & 1.479 & 0.262  & 0.7182 & 0.0083 \\
        $0.63 < z_\mathrm{ph} < 0.87$ & 1.484 &  0.259 & 0.6887 & 0.0126 \\
        $0.87 < z_\mathrm{ph} < 2.0$  & 1.461 & 0.301  & 0.6154 & 0.0145 \\
        \hline\hline
    \end{tabular}
    \label{tab:SourceSample}
\end{table}

\subsection{Log-normal realisations}
\label{sec:Simslog-normal}

In order to validate our covariance matrix and parameter inference pipeline, we rely on the simulation of a large ensemble of log-normal random fields.
The log-normal distribution has been used in a broad range of applications for modelling cosmic fields \citep{ColesandJones91}.
Their efficacy has been demonstrated in approximating the 1-point probability density function (PDF) of weak lensing convergence/shear \citep{2011A&A...536A..85H,flask} and the distribution of late-time matter density contrast fields \citep{Friedrich_DensitySplitStats18,Gruen_DensitySplitStatsDes18}. 
In the context of DES, \cite{Clerkin_LNDes17} further validated the log-normal distribution assumption using Science Verification data for the convergence field.
More recently, \cite{Friedrich_2021} employed it to compute and validate covariance matrices for the full combination of weak lensing and galaxy clustering correlation functions, the 3$\times$2pt data vector in the configuration space of DES Y3.

In this work we use the implementation of the Full-sky log-normal Astro-fields Simulation Kit \citep[\flask;][]{flask} to generate a suite of 1800 log-normal realisations of our data vector.

The \flask requires a set of angular power spectra as its primary input.
These spectra must include the auto- and cross-correlations of all the desired cosmic fields, simulated as \healpix maps.
We generate them using the theoretical framework presented in Section~\ref{sec:modelling} at the fiducial \redmagic + \metacalibration cosmology of Table \ref{tab:parameters}.
A more detailed description of the parameters of Table \ref{tab:parameters} is found in Section \ref{sec:modelling} and the power spectra measurements in the suite of log-normal realisations are presented in Appendix \ref{app:measurements}.

\begin{table*}
\fontsize{8pt}{8pt}\selectfont
\centering
\caption{
    The parameters used in the present analyses.
    We show the fiducial values used for the construction of the log-normal realisations and the prior probability distributions used for the Bayesian parameter inference. Priors for the \redmagic + \metacalibration analysis follow the ones described in \protect\cite{2x2_redmagic} while the priors for the \maglim + \metacalibration are the same as in \protect\cite{2x2_maglim}. Uniform priors are described by $\mathcal{U}(x,y)$, with $x$ and $y$ denoting the prior edges, and Gaussian priors are represented by $\mathcal{N}(\mu,\sigma)$, with $\mu$ and $\sigma$ being the mean and standard deviation. The dark energy equation of state $w$ is fixed at $-1$ for \LCDM chains, while for \wCDM it varies following the indicated uniform prior. The main analyses were performed only with linear galaxy bias in their modelling; the non-linear galaxy bias parameters were used in some additional runs (see Section~\ref{sec:scale cuts} and \ref{sec:robustness_tests}).
}
\label{tab:parameters}
\begin{tabular}{lcc|lcc}
\hline\hline
\multicolumn{3}{c}{\textbf{Cosmology}} & \multicolumn{3}{c}{\textbf{Intrinsic Alignments}} \\
\cmidrule(lr){1-3}\cmidrule(lr){4-6}
Parameters & Fiducial values & Priors   & Parameters & Fiducial values & Priors \\
\cmidrule(lr){1-3}\cmidrule(lr){4-6}
$\Omega_m$                  & 0.3       & $\mathcal{U}(0.1, 0.9)$ &     $a_1$               & 0.7                     &  $\mathcal{U}(-5.0, 5.0)$\\

$h_0$                      & 0.69      & $\mathcal{U}(0.55, 0.91)$ & $a_2$               & -1.36                   &  $\mathcal{U}(-5.0, 5.0)$\\

$\Omega_b$                  & 0.048     & $\mathcal{U}(0.03, 0.07)$ &   $\alpha_1$          & -1.7                    &  $\mathcal{U}(-5.0, 5.0)$\\

$n_s$            & 0.97      & $\mathcal{U}(0.87, 1.07)$ &  $\alpha_2$          & -2.5                    &  $\mathcal{U}(-5.0, 5.0)$\\

$A_s 10^{9}$ & 2.19      & $\mathcal{U}(0.5, 5.0)$ &    $b_{\text{TA}}$     & 1.0                     &  $\mathcal{U}(0.0, 2.0)$\\
$\Omega_\nu h^2$          & 0.00083   & $\mathcal{U}(0.0006, 0.00644)$ &    $z_0$               & 0.62                    &  Fixed\\
$w$ & $-1$ & \LCDM: fixed | \wCDM: $\mathcal{U}(-2, -0.33)$ & & & \\
\hline\hline
\multicolumn{3}{c}{\textbf{\redmagic + \metacalibration}} & \multicolumn{3}{c}{\textbf{\maglim + \metacalibration}} \\
\cmidrule(lr){1-3}\cmidrule(lr){4-6}
Parameters & Fiducial values & Priors   & Parameters & Fiducial values & Priors \\

\cmidrule(lr){1-3}\cmidrule(lr){4-6}
\multicolumn{3}{l}{Linear galaxy bias} & \multicolumn{3}{l}{Linear galaxy bias} \\
$b_i$              & 1.7, 1.7, 1.7, 2.0, 2.0 & $\mathcal{U}(0.8, 3.0)$
&$b_i$              & 1.5, 1.8, 1.8, 1.9, 2.3, 2.3 & $\mathcal{U}(0.8, 3.0)$\\

\cmidrule(lr){1-3}\cmidrule(lr){4-6}
\multicolumn{3}{l}{Non-linear galaxy bias} & \multicolumn{3}{l}{Non-linear galaxy bias} \\
$b^i_1 \sigma_8$              & 1.42, 1.42, 1.42, 1.68, 1.68 & $\mathcal{U}(0.67, 2.52)$
&$b_1^i \sigma_8$              & 1.43, 1.43, 1.43, 1.69, 1.69, 1.69 & $\mathcal{U}(0.67, 3.0)$\\

$b^i_2 \sigma_8$              & 0.16, 0.16, 0.16, 0.35, 0.35 & $\mathcal{U}(-3.5, 3.5)$
&$b_2^i \sigma_8$              & 0.16, 0.16, 0.16, 0.36, 0.36, 0.36 & $\mathcal{U}(-4.2, 4.2)$\\

\cmidrule(lr){1-3}\cmidrule(lr){4-6}
\multicolumn{3}{l}{Shear calibration} & \multicolumn{3}{l}{Shear calibration} \\
$m^1$              & $-0.0063 $                       & $\mathcal{N}(-0.0063, 0.0091)$
&$m^1$              & $-0.006    $                  & $\mathcal{N}(-0.006, 0.008)$\\
$m^2$              &$ -0.0198  $                      & $\mathcal{N}(-0.0198, 0.0078)$
&$m^2$              & $-0.010  $                       & $\mathcal{N}(-0.010, 0.013)$\\
$m^3$              & $-0.0241   $                      & $\mathcal{N}(-0.0241, 0.0076)$
&$m^3$              & $-0.026 $                       & $\mathcal{N}(-0.026, 0.009)$\\
$m^4$              &$ -0.0369  $                       & $\mathcal{N}(-0.0369, 0.0076)$
&$m^4$              & $-0.032$                        & $\mathcal{N}(-0.032, 0.012)$\\
\cmidrule(lr){1-3}\cmidrule(lr){4-6}

\multicolumn{3}{l}{Source photo-$z$} & \multicolumn{3}{l}{Source photo-$z$} \\
$\Delta z^1_s$     & 0.0                         & $\mathcal{N}(0.0, 0.018)$ 
& $\Delta z^1_s$     & 0.0 & $\mathcal{N}(0.0, 0.018)$\\
$\Delta z^2_s$     & 0.0                         & $\mathcal{N}(0.0, 0.015)$ 
&$\Delta z^2_s$     & 0.0                         & $\mathcal{N}(0.0, 0.013)$ \\
$\Delta z^3_s$     & 0.0                         & $\mathcal{N}(0.0, 0.011)$
&$\Delta z^3_s$     & 0.0                         & $\mathcal{N}(0.0, 0.006)$\\
$\Delta z^4_s$     & 0.0                         & $\mathcal{N}(0.0, 0.017)$
&$\Delta z^4_s$     & 0.0                         & $\mathcal{N}(0.0, 0.013)$\\
\cmidrule(lr){1-3}\cmidrule(lr){4-6}

\multicolumn{3}{l}{Lens photo-$z$} & \multicolumn{3}{l}{Lens photo-$z$ shift} \\
$\Delta z^1_l$     & $0.006  $                       & $\mathcal{N}(0.006, 0.004)$ 
& $\Delta z^1_l$     & $-0.009$ & $\mathcal{N}(-0.009, 0.007)$\\
$\Delta z^2_l$     & $0.001$                         & $\mathcal{N}(0.001, 0.003)$ 
&$\Delta z^2_l$     & $-0.035$                         & $\mathcal{N}(-0.035, 0.011)$ \\
$\Delta z^3_l$     & $0.004$                         & $\mathcal{N}(0.004, 0.003)$
&$\Delta z^3_l$     & $-0.005$                         & $\mathcal{N}(-0.005, 0.006)$\\
$\Delta z^4_l$     & $-0.002$                         & $\mathcal{N}(-0.002, 0.005)$
&$\Delta z^4_l$     & $-0.007$                         & $\mathcal{N}(-0.007, 0.006)$\\
$\Delta z^5_l$     & $-0.007 $                        & $\mathcal{N}(-0.007, 0.01)$
&$\Delta z^5_l$     & 0.002                         & $\mathcal{N}(0.002, 0.007)$\\
  -   &- & -
&$\Delta z^6_l$     & 0.002                         & $\mathcal{N}(0.002, 0.008)$\\
\cmidrule(lr){1-3}\cmidrule(lr){4-6}

\multicolumn{3}{l}{Lens photo-$z$ stretch} & \multicolumn{3}{l}{Lens photo-$z$ stretch} \\
$\sigma z^1_l$     & 1.0                         & Fixed
&$\sigma z^1_l$     & 0.975                         & $\mathcal{N}(0.975, 0.062)$\\
$\sigma z^2_l$     & 1.0                         & Fixed
&$\sigma z^2_l$     & 1.306                         & $\mathcal{N}(1.306, 0.093)$\\
$\sigma z^3_l$     & 1.0                         & Fixed
&$\sigma z^3_l$     & 0.870                         & $\mathcal{N}(0.870, 0.054)$\\
$\sigma z^4_l$     & 1.0                         & Fixed
&$\sigma z^4_l$     & 0.918                         & $\mathcal{N}(0.918, 0.051)$\\
$\sigma z^5_l$     & 1.23                         & $\mathcal{N}(1.23, 0.054)$
&$\sigma z^5_l$     & 1.080                         & $\mathcal{N}(1.080, 0.067)$\\
  -   & -& -
&$\sigma z^6_l$     & 0.845                         & $\mathcal{N}(0.845, 0.073)$\\

\hline \hline

\end{tabular}
\end{table*}

\subsection{Angular power spectra measurements}
\label{sec:measurements}

We estimate angular power spectra of galaxy clustering (GCL) and galaxy-galaxy lensing (GGL) using the so-called pseudo-$C_\ell$
(PCL) or MASTER method~\citep{1973ApJ...185..413P,2002ApJ...567....2H,2005MNRAS.360.1262B}, as implemented in the \namaster
code\footnote{\tt github.com/LSSTDESC/NaMaster}~\citep{Alonso_2019}.

We commence by constructing weighted tomographic cosmic shear maps as
\begin{equation}
    \vec{\gamma}_p = \frac{\sum_{i\in p} v^{(\gamma)}_i \hat{\vec{\gamma}}_i}{\sum_{i\in p} v^{(\gamma)}_i},
\end{equation}
where the index $i$ runs over galaxies in the catalogue, index $p$ runs over pixels in map, $\hat{\vec{\gamma}}_i=(\hat\gamma_{1,i}, \hat\gamma_{2,i})$ is the calibrated galaxy shear and $v^{(\gamma)}_i$ its associated weight.

Analogously, tomographic galaxy over-density maps are given by
\begin{equation}
    \delta_p = \frac{N_p \sum_{p'} w^{(\delta)}_{p'}}{w^{(\delta)}_p \sum_{p'} N_{p'}} - 1
\end{equation}
where $N_p = \sum_{i\in p} v^{(\delta)}_i$ gives the number of galaxies at a given pixel $p$, with $v_i$ the weight associated with the $i$-th galaxy as given by the systematics weights (see Section~\ref{sec:data}).
When working with simulated log-normal realisations, we assume $v_i = 1$.
Whereas, $w^{(\delta)}_p$ gives the effective fraction of the area covered by the survey at pixel $p$.

In addition to the cosmic shear and galaxy clustering signal maps, the pseudo-$C_\ell$ method relies on the use of an angular window function, also known as the mask.
Such a mask encodes the information of the partial-sky coverage of the observed signal and is used to deconvolve this effect on the estimated bandpowers.
For the cosmic shear maps, we use the inverse-variance weighting scheme as presented in \cite{2021JCAP...03..067N}, and construct tomographic mask maps as 

\begin{equation}
    w^{(\gamma)}_p = \sum_{i\in p} v^{(\gamma)}_i.
\end{equation}
We note this approach assigns the weighted number of galaxies map as mask, thus there are different masks for each source tomographic bin.

The discrete nature of galaxies introduces a shot-noise contribution to the auto-correlation spectra.
We estimate this noise analytically, as specified below, and subsequently, we subtract it from our power spectrum estimates.
For the case of galaxy clustering we assume this noise to be Poissonian, and estimate it analytically for the pseudo-spectra following \cite{Alonso_2019}, \cite{2020JCAP...03..044N}, and \cite{2021JCAP...10..030G} as
\begin{equation}
    \label{eq:noise_clustering}
    \tilde{N}^{(\delta)}_\ell = \left( \frac{\sum_p w^{(\delta)}_p}{N_{\rm pix}} \right) \frac{1}{n_{\rm eff}},
\end{equation}
where $N_{\rm pix}$ is the total number of pixels in our maps.
We note that this noise is constant over all multipoles.
For \healpix-based pixelization, this is given in terms of the \healpix resolution parameter \citep{2005ApJ...622..759G}, $N_{\rm side}$ as $N_{\rm pix} = 12 N_{\rm side}^2$.
Note the first term in parentheses is the mean of the mask on the sphere.
The second term is the Poissonian noise in terms of the effective mean angular number density, $n_{\rm eff}$, when accounting for the galaxy weights, $v^{\delta}_i$, as
\begin{equation}
    n_{\rm eff} = \frac{\left( \sum_i v^{(\delta)}_i \right)^2}{\Omega_{\rm pix} \sum_p w^{(\delta)}_p \sum_i \left( v^{(\delta)}_i \right)^2},
    \label{eq:WeightedAngularNumberDensity}
\end{equation}
where $\Omega_{\rm pix}$ is the area of each \healpix pixel at the given resolution. 

For the case of cosmic shear, the pseudo-spectra noise is computed following \cite{2020JCAP...03..044N} as
\begin{equation}
    \label{eq:noise_shear}
    \tilde{N}^{(\gamma)}_{\ell>2} = \Omega_{\rm pix} \frac{\sum_p \sum_{i\in p} \left(v^{(\gamma)}_i\right)^2 \sigma_{e,i}^2}{N_{\rm pix}},
\end{equation}
where $\sigma_{e, i}^2 = \left( \hat{\gamma}_{1,i}^2 + \hat{\gamma}_{2,i}^2 \right) / 2$ is the RMS noise per galaxy for a given shape estimator.

\begin{figure*}
  \centering
  \includegraphics[width=0.99\textwidth]{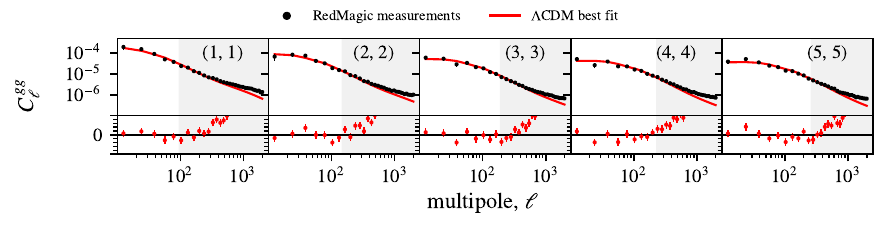}
  \centering
  \includegraphics[width=0.99\textwidth]{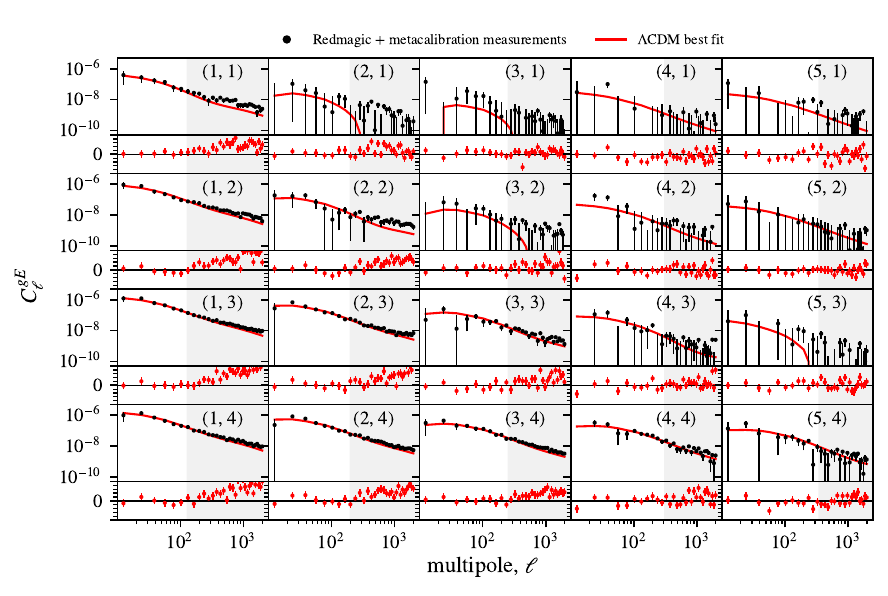}
  \caption{Angular power spectrum measurements of galaxy clustering ($C_\ell^{\text{gg}}$) and galaxy-galaxy lensing ($C_\ell^{\text{gE}}$) from the DES Y3 \redmagic and \metacalibration catalogues, along side with theoretical prediction calculated at the \LCDM best-fit values. Residual plots compare measurements with the theoretical prediction.}
  \label{fig:redmagic_measurements}
\end{figure*}
\begin{figure*}
  \centering
  \includegraphics[width=0.99\textwidth]{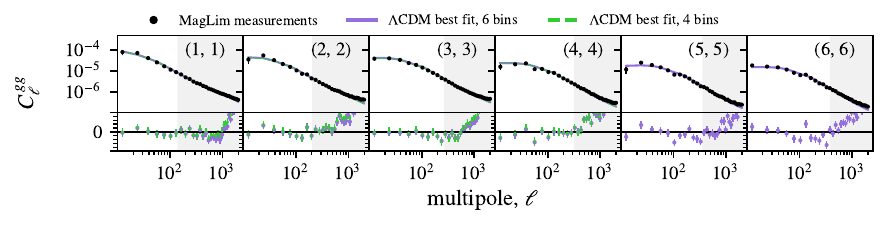}
  \centering
  \includegraphics[width=0.99\textwidth]{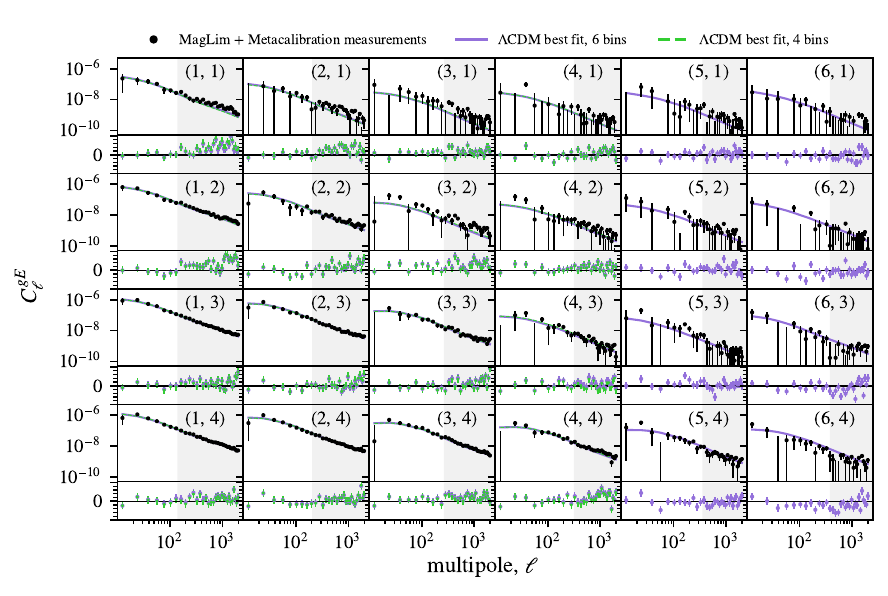}
  \caption{Angular power spectrum measurements of galaxy clustering ($C_\ell^{\text{gg}}$) and galaxy-galaxy lensing ($C_\ell^{\text{gE}}$) from the DES Y3 \maglim and \metacalibration catalogues, alongside with theoretical prediction calculated at the \LCDM best-fit values from the analysis using all six tomographic lenses (solid, purple line) and only the first four bins (dashed, green line). Residual plots compare measurements with the theoretical prediction from the six-bin (purple markers) and four-bin (green points) analysis. The indexes $(z_i, z_j)$ in each subplot indicate the redshift bin combination. For GGL, $z_i$ and $z_j$ refer to the bin of the lens and source, respectively.}
  \label{fig:maglim_measurements}
\end{figure*}

Finally, the different pseudo-spectra are binned into bandpowers.
The chosen bandpowers are a set of 32 square-root-spaced bins within the multipole interval $\ell = [8, 2048]$.
The angular power spectra measurements for galaxy clustering $C^{\text{gg}}_\ell$ and galaxy-galaxy lensing $C^{\text{gE}}_\ell$ using the \redmagic and \maglim lens samples are shown in Figures~\ref{fig:redmagic_measurements} and \ref{fig:maglim_measurements}, along with the theory prediction calculated at the \LCDM best-fit values. For \maglim, we use one prediction considering only the first four tomographic bins (green) and another using all six bins (purple). 
The residual plots comparing the measurements to these predictions are shown below for each redshift combination. The measurements of the 1800 log-normal realisations can be found in Appendix ~\ref{app:measurements}.

\section{Methodology}
\label{sec:methodology}

\subsection{Theoretical modelling }
\label{sec:modelling}

Our theoretical framework for modelling the angular power spectra draws upon the formalism presented in \cite{DesMethods_Krause21}.

We calculated the galaxy clustering angular power spectra using the full non-Limber approach detailed in Section III.B of \cite{DesMethods_Krause21},
\begin{equation}
    C^{gg}_{ij}(\ell) = \frac{2}{\pi} \int_0^\infty \frac{{\rm d} k}{k}\, k^3 P_{\rm NL}(k) \Delta_{\delta_{\rm g}}^i(k, \ell) \Delta_{\delta_{\rm g}}^j(k, \ell),
    \label{eq:theory_Cell_clustering}
\end{equation}
where the kernel accounts for linear density growth (D), redshift-space distortions (RSD) and lensing magnification ($\mu$), $\Delta^i_{\delta_{\rm g}}(k,\ell)=\Delta^i_{\rm D}(k,\ell)+\Delta^i_{\rm{RSD}}(k,\ell)+\Delta^i_{\mu}(k,\ell)$.
The specific form of each term is provided by \cite{DesMethods_Krause21} and \cite{nonlimber_fang20}.
Numerical integrations were performed using the FFTLog algorithm \citep{Hamilton_fftlog00} as implemented by \cite{nonlimber_fang20}.

The galaxy-galaxy lensing spectra, on the other hand, were evaluated under the Limber approximation,
\begin{align}
    C^{gE}_{ij}(\ell) = \int {\rm d}\chi \frac{q_{\delta_g}^i(\chi) q_\gamma^j(\chi)}{\chi^2} P_{\rm NL}\left(k = \frac{\ell+0.5}{\chi}, z(\chi)\right).
    \label{eq:theory_Cell}
\end{align}
In both Equations \eqref{eq:theory_Cell_clustering} and \eqref{eq:theory_Cell}, we refer to $\delta_{\rm g}$ as the \emph{lens galaxy sample} overdensity field and $\gamma$ as the \emph{source galaxy sample} shear field and $i, j$ run over tomographic redshift bins for the corresponding galaxy sample.

Furthermore, Equations \eqref{eq:theory_Cell_clustering} and \eqref{eq:theory_Cell} can be understood as projections of the nonlinear total matter power spectrum, $P_{\rm NL}(k, z)$, computed using CAMB \citep{CAMB_Lewis20} and \halofit \citep{halofit_Takahashi12}, by specific radial kernels per tomographic bin along the comoving distance $\chi$ or redshift $z$.
These kernels encode the response to the large-scale structure of the different probes at different scales and are given by
\begin{equation}
    q^i_{\delta_g}(\chi) = b^i \left( k, z(\chi) \right) n_{\delta_{g}}^i \left( z(\chi) \right) \frac{{\rm d} z}{{\rm d} \chi},
    \label{eq:delgag_kernel}
\end{equation}
\begin{equation}
    q^{i}_{\gamma}(\chi) = \frac{3 H_0^2 \Omega_{\rm m}}{2 c^2} \frac{\chi}{a(\chi)} \int_{\chi}^{\infty} {\rm d} \chi' n_{\gamma}^i \left( z(\chi') \right) \frac{{\rm d} z}{{\rm d} \chi'}  \frac{\chi' - \chi}{\chi'},
    \label{eq:gamma_kernel}
\end{equation}
where $H_0$ is the Hubble's constant, $\Omega_{\rm m}$ the matter density, $a(\chi)$ the scale factor corresponding to the comoving distance $\chi$, $b(k,z)$ the scale-dependent galaxy bias, and $n^i_{{\delta_{g}}/\gamma}(z)$ the normalised redshift distributions of the lens/source galaxies in the tomographic bin $i$.

We consider two models for the galaxy bias $b(k,z)$. 
The first, and fiducial choice, is a linear bias model where $b(k,z)=b^{i}$ is a constant free parameter for each tomographic bin $i$.
The second one is a nonlinear biasing model presented by \cite{nonlinearbias_saito14} and \cite{2x2_redmagic}, consisting of a perturbative galaxy bias
model to third order in the density field with four parameters: $b^{i}_1$ (linear bias), $b^{i}_2$ (local quadratic bias), $b^{i}_{s^2}$ (tidal quadratic bias) and $b^{i}_{\rm 3nl}$ (third-order non-local bias).
Following \cite{2x2_redmagic} and \cite{2x2_maglim}, we fix the bias parameters $b_{s^2}^i$ and $ b_{\rm 3nl}^i $ to their co-evolution value of $b_{s^2}^i=-4(b^i - 1)/7$ and $b_{\rm 3nl}^i=b^i - 1$ \citep{nonlinearbias_saito14}, making the total number of free parameters for this bias model two per tomographic bin $i$.
We further note that for this second model, the power spectra from \halofit is not used.
Instead, different kernels for each bias term are computed from the linear power spectrum, following \cite{nonlinearbias_saito14} and \cite{2x2_redmagic}.

To account for the contribution to the observed galaxy shapes caused by the gravitational tidal field, the so-called intrinsic alignment (IA) effect, we adopt the tidal alignment tidal torquing (TATT) model of \cite{tatt_blazek19}.
The TATT model has five parameters: $a_1$ and $\alpha_1$ characterise the amplitude and redshift dependence of the tidal alignment; $a_2$ and $\alpha_2$ characterise the amplitude and redshift dependence of the tidal torquing effect and $b_{\rm TA}$ accounts for the fact that our measurement is weighted by the observed galaxy counts.
Following \cite{2x2_redmagic} and \cite{2x2_maglim} we will also compare our results using a simpler IA model, the nonlinear alignment (NLA) model of \cite{nla_bridle07}.
The NLA model is equivalent to the TATT model in the limit where $a_2,\alpha_2,b_{\rm TA}\to0$, thus having two free parameters. 

Foreground structure can distort the observed lens galaxy properties due to gravitational lensing magnification effects.
Such distortions are commonly known as lens magnification, and impact both the apparent position and the distribution of light received from individual galaxy images.
We model the effect of lens magnification following \cite{DesMethods_Krause21} and \cite*{y3-2x2ptmagnification} by modifying the lens galaxy overdensity kernel, Equation \eqref{eq:delgag_kernel}, as
\begin{equation}
    q^i_{\delta_{\rm g}}(\chi) \rightarrow q^i_{\delta_{\rm g}}(\chi) (1+C^{i}_{\rm g}\kappa^{i}_g),
\end{equation}
where $\kappa_{\rm g}^i$ is the tomographic convergence field, as defined in \cite{DesMethods_Krause21} and \cite*{y3-2x2ptmagnification}, and $C_{\rm g}^i$ are the magnification bias coefficients.
We fix the values of $C_{\rm g}^i$ to the ones estimated by \cite*{y3-2x2ptmagnification} as listed in Table~\ref{tab:LensSamples}.

To account for possible residual uncertainty in both lens and source galaxies redshift distributions, we introduce shift parameters, $\Delta z^i$, when modelling  the redshift distributions,
\begin{equation}
    n^{i}(z) \rightarrow n^{i}(z-\Delta^{i}_{z}).
\end{equation}    
For the lens sample, motivated by \cite{clusteringz_cawthon22} and \cite{2x2_maglim} we additionally introduce stretch parameters ($\sigma^i_{z}$), as 
\begin{equation}
    n^{i}(z) \rightarrow \frac{n^{i}}{\sigma_{z}^{i}} \left( \frac{z-\langle z \rangle}{\sigma_{z}^{i}} + \langle z \rangle -\Delta^{i}_{z} \right),
\end{equation}
where $\langle z \rangle$ is the mean redshift of the $i$-th tomographic bin.

To account for possible residual uncertainty in the shear calibration, we introduce multiplicative factors to the shear kernel, Equation \eqref{eq:gamma_kernel}, as
\begin{equation}
    q^i_\gamma(\chi) \rightarrow  (1+m^{i}) q^i_{\gamma}(\chi) 
\end{equation}
where $m^{i}$ is the shear calibration bias for source bin $i$.

Finally, the theoretical angular power spectrum is binned into bandpowers.
This is done by filtering the predictions with a set of bandpower windows, $\mathcal{F}_{q\ell}^{ab}$, consistent with the pseudo-$C_\ell$ approach we follow for the data estimates (see Section~\ref{sec:measurements}).
Thus the final model for a bandpower, $\ell \in q$, is computed as
\begin{equation}
\label{eq:bpws-modelling}
    \mathbf{C}_{(i,j)}(q) = \sum_{\ell \in q} \mathcal{F}_{q \ell}^{(i,j)} \mathbf{C}_{(i,j)}(\ell)
\end{equation}
where $(i, j)$ represents the tomographic redshift bin pair, and a vector notation, $\mathbf{C} = \left( C^{gg}, C^{gE}, C^{gB} \right)$, is required to account for the $E/B$--mode decomposition of the shear field.
We refer the reader to \citet{Alonso_2019} for the detailed expressions for the bandpower windows and details about the $E/B$--mode decomposition.

All the different pieces for the modelling presented above are integrated as modules in the \cosmosis\footnote{\url{https://github.com/joezuntz/cosmosis}} framework \citep{Zuntz_2015} in an analogous way to what was done for the configuration-space analysis presented in \cite{2x2_maglim} and \cite{2x2_redmagic}.

The complete set of parameters of the theoretical modelling is summarised in Table \ref{tab:parameters}, including their fiducial values and priors. For \LCDM analyses, we sample over the matter density $\Omega_m$, the Hubble parameter $h_0$, the amplitude of primordial scalar
density fluctuations $A_s$, the spectral index $n_s$, the baryonic density $\Omega_b$, and the massive neutrino density $\Omega_\nu$. The equation of state of dark energy $w$ is set as a free parameter in the \wCDM analyses. Following the DES standard, we quote our results in terms of the clustering amplitude, defined as 
\begin{equation}
    S_8 \equiv \sigma_8 \left( \frac{\Omega_m}{0.3}  \right)^{0.5},
\end{equation}
where $\sigma_8$ is the amplitude of mass fluctuations on 8 Mpc$/h$ scale in linear theory.

\subsection{Scale cuts}
\label{sec:scale cuts}

\begin{figure}
\includegraphics[width=\linewidth]{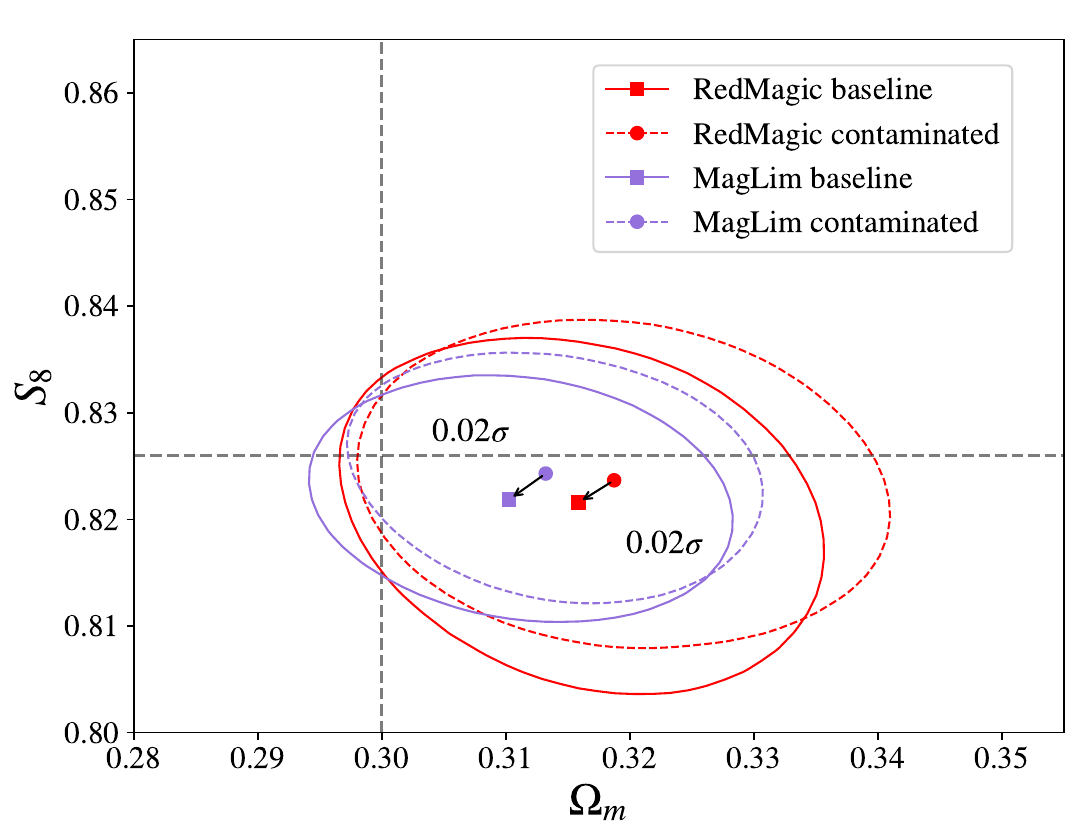}\par
\caption{2-dimensional posterior of $S_8$ and $\Omega_m$ for \redmagic and \maglim baseline and contaminated data vectors under the extended (8, 6) Mpc/$h$ scale cut, both analysed with linear bias and \halofit model. 
The ellipses represent $0.3\sigma_{\text{2D}}$ contours around their best-fit values marked in the centre. The arrows and numbers show the distance between the best-fit of baseline and contaminated chains.}
\label{fig:0.3sigma}
\end{figure}

Due to the fact that the chosen theoretical modelling is not a complete astrophysical modelling $-$ lacking known effects that appear in smaller, non-linear scales such as baryonic dynamics and higher order non-linear galaxy bias $-$ using the full range of scales in this analysis would result in inaccurate results.
For that reason, finding the scales for which our modelling correctly predicts the physics involved is a crucial part of this work.

In order to find this range of scales, we test our pipeline with two simulated data vectors. One represents our fiducial modelling, while the other includes additional modelling for non-linear galaxy bias and baryons, the dominant non-linear effects of galaxy clustering and GGL.
From now on, they will be referred to as baseline and contaminated data vectors, respectively. The non-linear galaxy bias were modelled by including contributions from the local quadratic bias $b_2^i$ for every photo-$z$ bin of lenses $i$ (see Section \ref{sec:modelling}). Their fiducial values are the same ones from \citep{2x2_redmagic} and \citep{2x2_maglim}, and can be found in Table \ref{tab:parameters}.
The impact of the baryonic physics were modelled using hydrodynamic simulations with strong AGN feedback. In particular, we use the  AGN simulation \citep{vanDaalen2011} from the OverWhelmingly Large Simulations project (OWLS) suite \citep{Schaye2010}.
Following the approach of the DES Y3 methods paper \citep{DesMethods_Krause21}, we expect a valid scale cut to satisfy the criteria of $0.3\sigma_{\text{2D}}$ of compatibility in the $S_8 \times \Omega_m$ plane of posteriors, calculated at the best-fit values from both data vectors.

The galaxy-galaxy lensing signal in configuration space is inherently non-local.
This means the predicted signal at a given source-lens separation depends on the modelling of all scales within that separation, including non-linear small scales.
Several approaches have emerged to address the non-locality of the configuration space galaxy-galaxy lensing signal \citep{Baldauf2010,MacCrann2020,Park2021}, see \cite{PratZacharegkas2023} for a recent comparison of different approaches.
In this work, we do not implement any additional methodology to circumvent non-localities in GGL. 
In harmonic space, the modelling of a given scale does not exhibit explicit dependence on smaller scales.
Nevertheless, we try to be extra careful about the scales we are including in the analysis.
We present two approaches: the \textit{conservative} and \textit{extended} scale cuts. For both, we made sure that the baseline and contaminated data vectors have a compatibility well under the $0.3\sigma_{\text{2D}}$ criterion. 
We evaluated compatibility using our standard pipeline with the priors of Table \ref{tab:parameters}.
To remain conservative and apply more stringent tests to the models considered, we did not perform nuisance parameter marginalisation, following the prescription in \cite{DesMethods_Krause21}.
The results are summarised in Figure \ref{fig:0.3sigma}.
The ellipses represent the marginalised $0.3\sigma_{\text{2D}}$ contours of the baseline and contaminated data vectors under the extended scale cut, and the difference between the peaks of these posteriors are denoted by the arrows. 

To select these scale cuts, we first defined a minimum physical distance $R_{\text{min}}$ that is associated to a maximum comoving Fourier mode $k_{\text{max}} = 1/R_{\text{min}}$. The maximum multipole $\ell_{\text{max}}$ can then be extracted for every redshift bin of lenses by the following relation:
\begin{equation}
    \ell_{\text{max}} = k_\text{max} \, \chi(\langle z_i \rangle)
    \label{eq:lmax}
\end{equation}
where $\langle z_i \rangle = \left( \int z \, n_i(z) \, dz \right) / \left( \int n_i(z) \, dz \right) $ is the mean redshift for the $i$th tomographic set of lenses and $\chi$ is the comoving distance assuming the fiducial cosmology of the analysis (see Table \ref{tab:parameters}). 
This means that GGL combinations sharing the same lenses also share the same $\ell_{\text{max}}$.

From this relation come our two sets of scale cuts:
the \textit{conservative} approach using $R_{\text{min}}^{\text{gcl}} = 8 \text{ Mpc}/h$ and $R_{\text{min}}^{\text{ggl}} = 12 \text{ Mpc}/h$ for clustering and GGL respectively; and the \textit{extended} approach, that goes to lower scales in GGL with $R_{\text{min}}^{\text{ggl}} = 6 \text{ Mpc}/h$.
Both methodologies follow the criteria from DES configuration space analyses. The $R_{\text{min}}^{\text{ggl}}$ reference in the conservative approach is the same as the one in \cite{desy1_3x2pt}, when the GGL data also did not receive special modelling for its non-localities. On the other hand, the minimum physical scale of $R_{\text{min}}^{\text{ggl}} = 6 \text{ Mpc}/h$ was chosen in \citep{Abbott_2022, Prat_2022_GGL, 2x2_redmagic}, when those non-localities were modelled using the point-mass marginalisation technique. Appendix \ref{app:scale_cuts} summarises the maximum multipoles associated with each one of these minimum distances for clustering and GGL.

\subsection{Likelihood}
\label{sec:likelihood}

We assume our power spectra measurements follow a multivariate Gaussian likelihood distribution with a fixed covariance matrix,

\begin{equation}
    -2 \ln \mathcal{L} \left( \mathbf{D} \right| \left. \Theta, M \right) = \chi^2 = \left[ \mathbf{D} - \mathbf{d}(\Theta) \right] \cdot \mathbf{C}^{-1} \cdot \left[ \mathbf{D} - \mathbf{d}(\Theta) \right],
    \label{eqn:likelihood}
\end{equation}
where $\mathbf{d}(\Theta)$ is the theoretical prediction for our data vector,
constructed by stacking the power spectra bandpowers for both probes, GCL and GGL, 
given the parameters $\Theta$ as described in Section \ref{sec:methodology}, and assuming a model $M$.

The corresponding measured data vector, $\mathbf{D}$, is also constructed by stacking the measured power spectra bandpowers for both probes over the different pair combinations of tomographic bins considered, 
accounting for scale cuts (see Sections \ref{sec:measurements} and \ref{sec:scale cuts}). 
We note that including shear ratios (SR) measurements is out of the scope of this work.
The SR methodology, as described by \cite*{sanchez22_desy3sr}, consists in taking the ratios of two GGL measurements that share the same lens tomographic bin. Under the limit that the lens distribution is sufficiently thin, this ratio loses its dependency on the power spectra and, thus, results in a geometrical measurement of the lensing system.
One can then use the SR from the small scales measurements of GGL, which would be discarded after the scale cut, to increase the constraining power of the analysis by improving the constraints of the systematics and nuisance parameters of the model.
These measurements were incorporated at the likelihood level in other DES Y3 works \citep{Abbott_2022, 2x2_redmagic, 2x2_maglim, Doux_2022}.
The methods to measure and apply shear ratios to harmonic space analyses are currently under development and will be implemented in future projects 
using DES Y6 data.

Finally, $\mathbf{C}$ is the joint covariance of galaxy clustering and galaxy-galaxy lensing power spectra.
It is analytically decomposed into Gaussian and non-Gaussian components arising from the cosmic shear and galaxy overdensity fields.
The Gaussian contribution is computed using \namaster to account for binning and mode coupling coming from partial sky coverage with the improved narrow-kernel approximation (iNKA) developed by \cite{2021JCAP...10..030G}, as optimised by \cite{2021JCAP...03..067N}.
We also account for the noise contribution to the covariance in the Gaussian term, using the analytical estimate from Equations \eqref{eq:noise_clustering} and \eqref{eq:noise_shear} as described in \cite{2021JCAP...03..067N}.
For this, we rely on the iNKA implementation of the general covariance calculator interface to be used for the Vera C. Rubin Observatory’s Legacy Survey of Space and Time (LSST\footnote{\url{https://www.lsst.org/lsst}}) Dark Energy Science Collaboration (DESC) \cite{lsst-desc}, \tjpcov\footnote{\url{https://github.com/LSSTDESC/TJPCov}}.

The non-Gaussian contribution comprises two components: 
i) the connected four-point covariance (cNG), arising from the joint cosmic shear and galaxy clustering trispectra; and 
ii) the super-sample covariance (SSC), which accounts for correlations between Fourier modes used in the analysis and super-survey modes \citep{Takada_2013}.
The computation of both components utilises the implementation provided by \cosmolike \citep{Krause_2017,2020MNRAS.497.2699F}.
This follows the methodology outlined in \cite{DesMethods_Krause21}, which in turn draws upon formulae established in the works of \cite{Takada_2009} and \cite{Schaan_2014}.
We simplify the treatment of partial sky coverage and binning for the non-Gaussian contribution by scaling it with the observed fraction of the sky, $f_{\rm sky}$ and computing it on the grid of points defined by the effective multipole of each bandpower considered.

\begin{figure*}
    \centering
    \begin{multicols}{2}
    \includegraphics[width=1.35\linewidth]{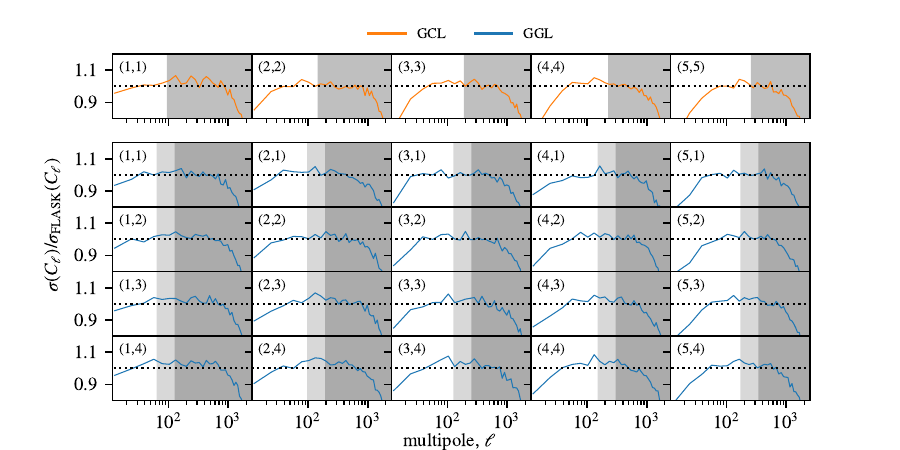}\par
    \includegraphics[width=0.75\linewidth]{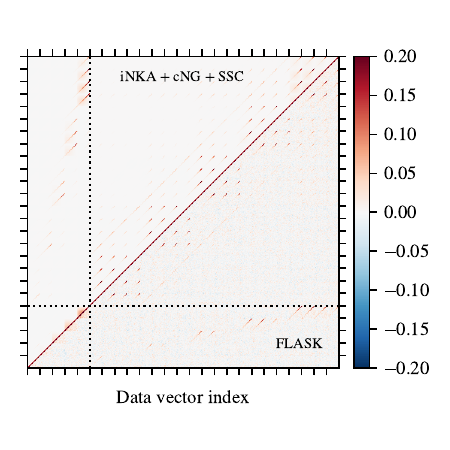}\par
    \end{multicols}
    \caption{
        The analytical covariance matrix used in this work is computed using \tjpcov, \namaster, and \cosmolike.
        Left: Comparison of the error bars estimated from log-normal realisations with the fiducial ones. We show the galaxy clustering (GCL) and galaxy-galaxy lensing (GGL) separately and the indexes $(z_i, z_j)$ in each subplot indicate the redshift bin combination. For GGL, $z_i$ and $z_j$ refer to the bin of the lens and source, respectively.
        Right: The full correlation matrix, from the log-normal realisations in the lower triangle, and the full analytical model including Gaussian (iNKA) and non-Gaussian (nNG + SSC) contributions in the upper triangle (note the normalisation in the range -0.2 to 0.2). 
    }
    \label{fig:covs}
\end{figure*}

A sample covariance was also computed from a set of 1800 log-normal simulations, and its comparison with our theoretical covariance is discussed in Section \ref{sec:validation_with_simulation}. The right-hand panel of Figure~\ref{fig:covs} shows both covariances side by side, while the left-hand panel shows the ratio of their diagonal. The compatibility between them is discussed in Section \ref{sec:validation_with_simulation}.

As discussed in \cite{Friedrich_2021}, calculating covariance matrices at a set of values considerably different from the best-fit values of the analysis can have a meaningful impact on the likelihood contours.
For that reason, the results on data shown in Section \ref{sec:results} were run twice, where in the second run the covariance was recalculated at the best-fit values of the first iteration.
Appendix \ref{app:cov_update} discusses the impact of updating the covariance in our main analyses.

The likelihood (Equation \eqref{eqn:likelihood}) is related to the posterior distributions of the parameters via the Bayes' theorem:
\begin{equation}
    P(\Theta|\mathbf{D} , M) \propto \mathcal{L}(\mathbf{D} | \Theta, M)\Pi (\Theta|M),
\end{equation}
where $\Pi (\Theta|M)$ is the prior probability distribution given a model $M$. The parameters constraints are reported by the mean of their marginalised posterior distributions and the $68\%$ confidence limits (C.L.) relative to this mean as error bars. The constraining power of different analyses are compared through the 2D Figure of Merit, defined as $\text{FoM}_{\Theta_1, \Theta_2} = (\text{det Cov}(\Theta_1, \Theta_2))^{-1/2}$, while the distance between their constraints is calculated as the difference between their best-fit values in terms of $\sigma_\text{2D}$, the marginalised $68\%$ C.L. on the $S_8 \times \Omega_m$ plane. The parameter inference was performed with the PolyChord sampler \citep{Handley_2015, Handley_2015_b}.


\subsection{Validations on simulated data}
\label{sec:validation_with_simulation}

\begin{figure}
  \centering
  \includegraphics[width=1\columnwidth]{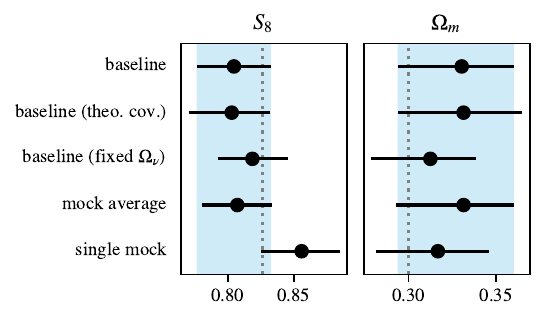}
  \caption{
      Comparison of marginalised constraints of \redmagic-like analyses between a fiducial, noiseless data vector (baseline), and data vectors constructed from the mean over the 1800 log-normal realisations (mock average) and a single realisation (single mock).
      First two rows: constraints of the baseline data vector with the sample covariance and the final theoretical covariance used in the \redmagic analysis. 
      Third row: constraint of the baseline data vector with the sample covariance, but fixing the value of $\Omega_\nu$.
      Forth and fifth rows: constraints of the mock average and single mock data vectors with the sample covariance.
      The vertical dotted lines represent the fiducial values used to construct both the noiseless data vector and the log-normal realisations. Shaded vertical regions are the $68\%$ C.L. marginalised regions from the constraints in the first row.
  }
  \label{fig:simulations}
\end{figure}

Before any work on data was done, all the aspects of the pipeline were extensively tested using simulated data vectors and covariances, to make sure that the developed methodology would not rely on biased expectations for the real measurements. A set of 1800 independent log-normal realisations generated by the code \flask \citep{flask} were produced for this purpose (Section \ref{sec:Simslog-normal}).
The measurements of the angular power spectra of this set are discussed in Appendix~\ref{app:measurements}.

The simulated covariance is a sample covariance matrix calculated out of this set of log-normal realisations. The right-hand panel of Figure \ref{fig:covs} shows the analytical and sample covariance side by side. In the left-hand panel, we compare the diagonal terms of both covariances, finding a good agreement in the valid range of scales after performing the scale cuts (non-shaded areas). The largest deviations occur at large scales $-$ more specifically, at the first and second bandpowers $-$ where the sample covariance has error bars around $\sim 10\%$ larger than the analytical for most combinations. We assessed the compatibility between our analytical and sample covariance by verifying their impact on cosmological constraints. Running the pipeline on our baseline data vector with the sample covariance and the final, analytical covariance for the \redmagic analysis showed an excellent agreement, resulting in a difference lower than $0.01\sigma_{\text{2D}}$. The marginalised constraints are shown in the first (sample cov.) and second (theoretical cov.) rows of Figure \ref{fig:simulations}.

One can notice that, despite the baseline angular power spectra being a noiseless, simulated data vector, the constraints found are not centred at their fiducial values $-$ although they are still consistent within $1\sigma$.
This deviation can be attributed
to the so-called prior projection effects.
In this particular analysis, the projection of the $\Omega_{\nu}$ prior is the main source of this effect (see \citealp{DesMethods_Krause21} for more discussions on that).
This statement is illustrated by the constraints shown in the third row of Figure \ref{fig:simulations}, where we run the pipeline on the same baseline data vector but with this parameter fixed, resulting in a much smaller deviation. This happens because $\Omega_{\nu}$ is prior dominated and its true value is set close to the boundary of the flat prior, which allows the peak of the likelihood to shift in the direction of its degeneracies. 

For the remaining two rows of Figure \ref{fig:simulations}, we perform some more pipeline tests with the sample covariance but using the average of all 1800 log-normal realisations (mock average) and a single, arbitrary realisation (single mock) as data vectors.
As expected, we see that the constraints obtained by the mock average case are consistent with the baseline ones, and that the single mock constraints are compatible with the true values within $1\sigma$.

Finally, taking advantage of the single mock data vector, we also check the goodness-of-fit compatibility between our analytical and sample covariance. After running the pipeline on the single mock with the sample covariance matrix, we obtained $\chi^2_F = 182$ at the best-fit parameters for the extended scale cut, with $227 - 30 = 197$ degrees of freedom (reduced $\chi^2_F = 0.92$), while the same test using our theoretical matrix resulted in $\chi^2_T = 208$ (reduced $\chi^2_T = 1.05$). 

This $\chi^2$ difference, $\Delta \chi^2_{T - F} = 11$, can be compared with analytical approximations for its expected value and variance (see Section~7 in \citealt{Andrade_Oliveira_2021} and \citealt{2020MNRAS.497.2699F}):
\begin{align}
    \text{E}[\Delta \chi^2_{T - F}] &= \text{Tr} ( \textbf{C}^{-1}_T  \textbf{C}_F) - N_D, \\
    \text{Var}[\Delta \chi^2_{T - F}] &= 2N_D + 2  \text{Tr} ( \textbf{C}^{-1}_T  \textbf{C}_F  \textbf{C}^{-1}_T  \textbf{C}_F) - 4  \text{Tr} ( \textbf{C}^{-1}_T  \textbf{C}_F),
\end{align}

where Tr represents the trace operator, $\textbf{C}_T$ and $\textbf{C}_F$ are the theoretical and the sample covariance from \flask, respectively, and $N_D$ is the number of degrees of freedom. These give us the estimation $\text{E}[\Delta \chi^2_{T - F}] = 9.11 \pm 9.00$, within a $1\sigma$ agreement with what was previously calculated for the particular realisation.


\section{Results}
\label{sec:results}

\begin{table*} 
  \caption{68\% C.L. marginalised cosmological constraints in \LCDM and \wCDM  using the combination of DES Y3 galaxy clustering and galaxy-galaxy lensing measurements (2$\times$2pt) in harmonic space (HS). In parenthesis after the constraints are the improvements of the error bars when compared to the configuration space (CS) results from \protect\cite{2x2_redmagic} and \protect\cite{2x2_maglim}. Last column shows the agreement, calculated as described in Section \ref{sec:likelihood}. 
  }
  \label{tab:results}
  \begin{tabular}{lcccccc} 
    \hline \hline
    \redmagic model & Scale cut & $\Omega_m$ & $S_8$ &$ 
    \sigma_8$ & $w$ & HS-CS agreement \\ 
    \hline \hline
    \LCDM & (8,6) Mpc/$h$ & $0.325\pm 0.040$ ($-19\%$) & $0.704\pm 0.029$ ($+2\%$) & $0.681^{+0.045}_{-0.072}$ & - &  $1.03 \sigma_{\text{2D}}$ \\
    
    \LCDM & (8,12) Mpc/$h$ & $0.328\pm 0.041$ ($-22\%$) & $0.690\pm 0.034$ ($-15\%$) & $0.665^{+0.041}_{-0.071}$ & - & $0.29 \sigma_{\text{2D}}$ \\
    
    \wCDM & (8,6) Mpc/$h$ & $0.301^{+0.037}_{-0.048}$ ($+7\%$) & $0.682^{+0.025}_{-0.039}$ ($0\%$) & $0.685^{+0.042}_{-0.052}$ & $-1.28\pm 0.29$ ($+8\%$) &  $0.95 \sigma_{\text{2D}}$  \\
    
    \hline \hline
    \maglim model & Scale cut & $\Omega_m$ & $S_8$ &$ 
    \sigma_8$ & $w$ & HS-CS agreement \\ 
    \hline \hline
    \LCDM, 6 bins & (8,6) Mpc/$h$ & $0.307^{+0.027}_{-0.037}$ ($+14\%$)& $0.753\pm 0.024$ ($+29\%$) & $0.748\pm 0.054$ & - & $0.9 \sigma_{\text{2D}}$\\

    \LCDM, 6 bins & (8,12) Mpc/$h$ & $0.315^{+0.028}_{-0.042}$ ($+9\%$) & $0.739\pm 0.033$ ($+3\%$) & $0.726\pm 0.061$ & -  & $1.27 \sigma_{\text{2D}}$ \\
    
    \wCDM, 6 bins & (8,6) Mpc/$h$ & $0.302\pm 0.036$ ($+17\%$) & $0.759\pm 0.032$ ($+20\%$) & $0.760\pm 0.051$ & $-1.01^{+0.24}_{-0.18}$ ($+2\%$) & $0.35 \sigma_{\text{2D}}$ \\

    \hline

    \LCDM, 4 bins & (8,6) Mpc/$h$ & $0.324^{+0.032}_{-0.047}$ ($-7\%$) & $0.779\pm 0.028$ ($+18\%$) & $0.754\pm 0.064$ & -  & $0.02 \sigma_{\text{2D}}$\\

    \LCDM, 4 bins & (8,12) Mpc/$h$ & $0.330^{+0.038}_{-0.043} $ ($-9\%$)  & $0.761\pm 0.035  $ ($-3\%$) & $0.731^{+0.058}_{-0.076}$ & -  & $0.14 \sigma_{\text{2D}}$\\

    \wCDM, 4 bins & (8,6) Mpc/$h$ & $0.321^{+0.040}_{-0.047}$ ($+17\%$) & $0.745\pm 0.039 $ ($+24\%$) & $0.726^{+0.057}_{-0.067} $  & $-1.26^{+0.34}_{-0.27}$ ($+7\%$)  & 0.42 $\sigma_{\text{2D}}$ \\

    \hline
  \end{tabular}
\end{table*}

After all the processes described in Section~\ref{sec:methodology} to validate our methodology with simulated data vectors and covariance, the analysis pipeline was applied to the real data. 
Due to the fact that the DES Y3 catalogues are public already, no specific blinding method was applied. That being said, as described in Section~\ref{sec:methodology}, every step of the pipeline had been previously tested and validated using simulations so that measurements of the catalogues and the main cosmological chains had to be run only once.

The main results of this work are presented in this Section. First, the constraints for $\Lambda$CDM (Section ~\ref{sec:lcdm}) and $w$CDM (Section ~\ref{sec:wcdm}) modelling are shown and discussed alongside the main results from \cite{2x2_redmagic} and \cite{2x2_maglim}. Table \ref{tab:results} summarises the constraints obtained, as well as the gain in constraining power in the 1D marginalised posteriors of $\Omega_m$, $S_8$ and $w$, and the agreement between harmonic and configuration space chains in the $S_8 \times \Omega_m$ 2D plane. In Appendix \ref{app:chi2} we discuss the goodness-of-fit of these results. Subsequently, a series of internal consistency and robustness tests are discussed (Section~\ref{sec:robustness_tests}).

As discussed in Section \ref{sec:likelihood}, after running the pipeline on data once, we recalculated the theoretical covariance at its best-fit values and ran the pipeline again with the updated covariance. Although the results of all these first iteration chains had a satisfactory goodness-of-fit, the runs for \maglim using all six photometric bins of lenses resulted in a $p-$value very close to the 0.01 requirement on goodness-of-fit used in the DES Y3 unblinding process (see Table \ref{tab:chi2}). This worst fit for the six-bin case after updating the covariance also happened in \cite{2x2_maglim}, where they localised the problem to be related to the last two redshift bins. Combining this with the fact that the best-fit values found for the six-bin case are considerably more sensitive to the covariance update, as discussed in Appendix \ref{app:cov_update}, we decided to quote the four-bin \maglim results as the fiducial ones for this work. For completeness, however, the results found using all six bins of \maglim are still shown in this Section. 

We note that there are other DES works that performed cosmological analyses using all six bins of \maglim in configuration space. In \cite{Giannini_2023} an alternative calibration for the 
\maglim redshift distributions was presented, and in \citet*{y3-2x2ptmagnification} the impact of the magnification bias was studied. 
We further note the results shown in these analyses are consistent with both the fiducial configuration space analysis \citep{2x2_maglim} and the present work.

\subsection{\boldmath{$\Lambda$}CDM}
\label{sec:lcdm}

\begin{figure*}
\begin{multicols}{2}
\includegraphics[width=1\linewidth]{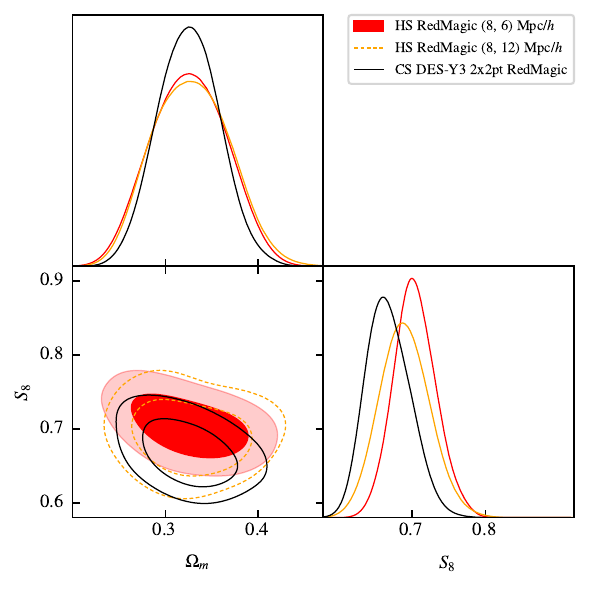}\par
\includegraphics[width=1\linewidth]{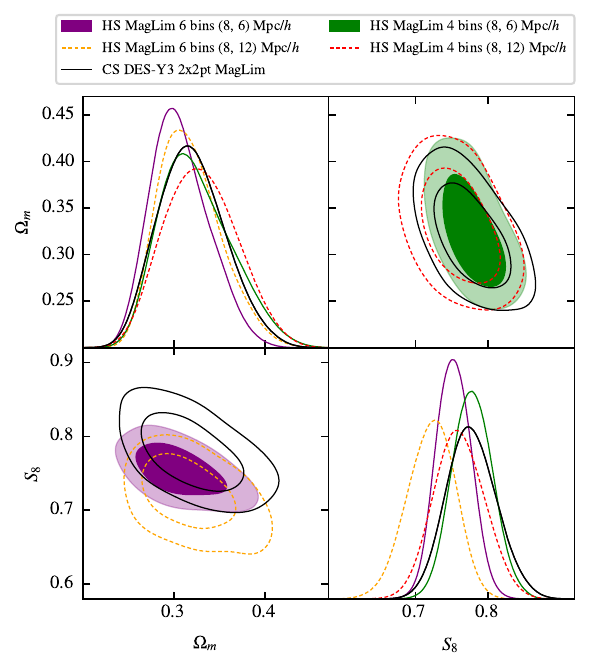}\par
\end{multicols}
\caption{marginalised constraints on $S_8$ and $\Omega_m$ in $\Lambda$CDM from galaxy clustering and galaxy-galaxy lensing joint analyses.
    Left: Harmonic (HS) and configuration space (CS) \redmagic constraints. 
    Right: Harmonic and configuration space \maglim  constraints. 
    Configuration space curves are the main results from \protect\cite{2x2_redmagic} and \protect\cite{2x2_maglim}, which includes shear ratio measurements in their analysis and 4 tomographic bins of lenses for \maglim.
}
\label{fig:chains}
\end{figure*}

The constraints derived from the harmonic space (HS) pipeline developed for this study, using galaxy clustering and galaxy-galaxy lensing measurements in $\Lambda$CDM analyses, are presented in Figure~\ref{fig:chains} for \redmagic (left-hand panel) and \maglim (right-hand panel; HS contours in the lower triangle include all six tomographic bins, while only the first four are included in the upper triangle) samples. The results for the DES Y3 2$\times$2pt configuration space analyses \citep{2x2_redmagic, 2x2_maglim} are also shown for comparison. It is important to emphasise that even though the harmonic space scale cuts were based on the same physical scale threshold as the real space ones, that does not mean both analyses were performed using the exact same information. Furthermore, these configuration space curves use their fiducial analysis choice, which includes shear ratios measurements. As mentioned in Section \ref{sec:likelihood}, the harmonic space curves do not include shear ratios.
The marginalised posterior probability distributions for $\Omega_m$, $S_8$ and $\sigma_8$ are summarised in Table \ref{tab:results}.

From Figure~\ref{fig:chains}, one can immediately notice that the \redmagic constraints point to lower values of $S_8$, both in harmonic and configuration space. This is properly discussed in \citep{2x2_redmagic}, where they found a bias in the galaxy selection of the fiducial sample of \redmagic. This issue was responsible for the best-fit values of the galaxy bias $b_i$ from the clustering part of the data vector to be systematically higher than the galaxy-galaxy lensing part, thus implying that the phenomenological parameter $X_\text{lens}^i = b^i_{\text{GGL}}/b^i_{\text{GCL}}$, often referenced as ``de-correlation'', would not be equal to 1 as predicted from local biasing models. In fact, they found $X_\text{lens} = 0.9\pm 0.03$ for the fiducial \redmagic catalogue when using a single de-correlation parameter for all redshift bins $-$ a $3.5\sigma$ deviation from 1. They were able to trace back the source of this de-correlation to be associated with the goodness-of-fit threshold $\chi^2_{\text{RM}}$ for a galaxy to be classified as part of the \redmagic sample in the procedure described in \citep{sv-redmagic}. Creating a new \redmagic sample with a broader $\chi^2_{\text{RM}}$ was enough to recover $X_\text{lens}$ compatible with 1.

\begin{figure}
  \centering
  \includegraphics[width=0.99\linewidth]{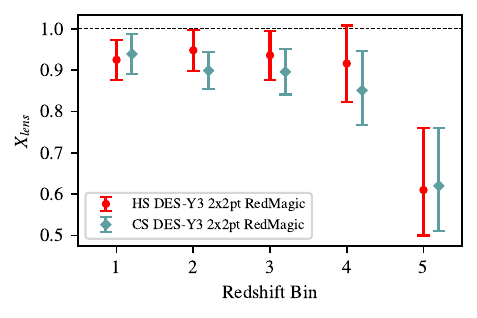}
  \caption{
    Constraints on the de-correlation parameter $X_{\text{lens}}$ for each tomographic redshift bin of the fiducial \redmagic sample. Harmonic space results (HS) are represented by the red, circle markers, while the green, diamond markers represents configuration space (CS) from \protect\cite{2x2_redmagic}.
  }
  \label{fig:xlens}
\end{figure}

In this work we only perform analyses using the fiducial \redmagic sample. Although the harmonic space constraints of this sample also point to lower values of $S_8$ when compared to \maglim results, we notice that these constraints are not as low as the one found in configuration space, resulting in a tension of $3.01\sigma_{\text{2D}}$ and $2.01\sigma_{\text{2D}}$ between \redmagic and \maglim four and six bins in HS, respectively, under the extended scale cut. The de-correlation parameter $X_{\text{lens}}$ for the \redmagic fiducial sample in harmonic space is consistent with what was found in configuration space, although slightly more compatible with 1 due to the higher $S_8$, as indicates Figure \ref{fig:xlens}. In order to compare the $X_{\text{lens}}$ constraints from HS and CS, we followed the prescription in \cite{2x2_redmagic}, where the cosmological parameters were fixed at the DES Y1 best-fit values \citep{desy1_3x2pt}.

From Figure~\ref{fig:chains} and Table~\ref{tab:results} 
we can see that the harmonic space results using the extended scale cut (shaded curves) have tighter constraints for the clustering amplitude $S_8$ in all three analyses (\redmagic and \maglim four and six bins) when compared to the configuration space ones (black lines). 
On the other hand, while we do find tighter constraints for matter energy density $\Omega_m$ in the four and six bins analyses of \maglim, the constraining power of \redmagic for this parameter is weaker in harmonic space. The \LCDM 4 and six bins analyses of \maglim had an improvement in FoM of $64\%$ and $14\%$ for our extended scale cut, while \redmagic had a decrease of $-15\%$.

The level of agreement between the harmonic and configuration space results can be quantified on the $S_8 \times \Omega_m$ plane, as shown in the last column of Table \ref{tab:results}. For all \LCDM cases we see deviations lower than $1\sigma$, except for the $1.03\sigma$ and $1.27\sigma$ found for \redmagic with the extended and \maglim six bins with the conservative scale cut, which are still statistically compatible with the configuration space results. In particular, the \maglim results with four redshift bins shows a strong agreement with CS results ($0.02\sigma_{\text{2D}}$ and $0.14\sigma_{\text{2D}}$). 

\begin{figure}
  \centering
  \includegraphics[width=0.99\linewidth]{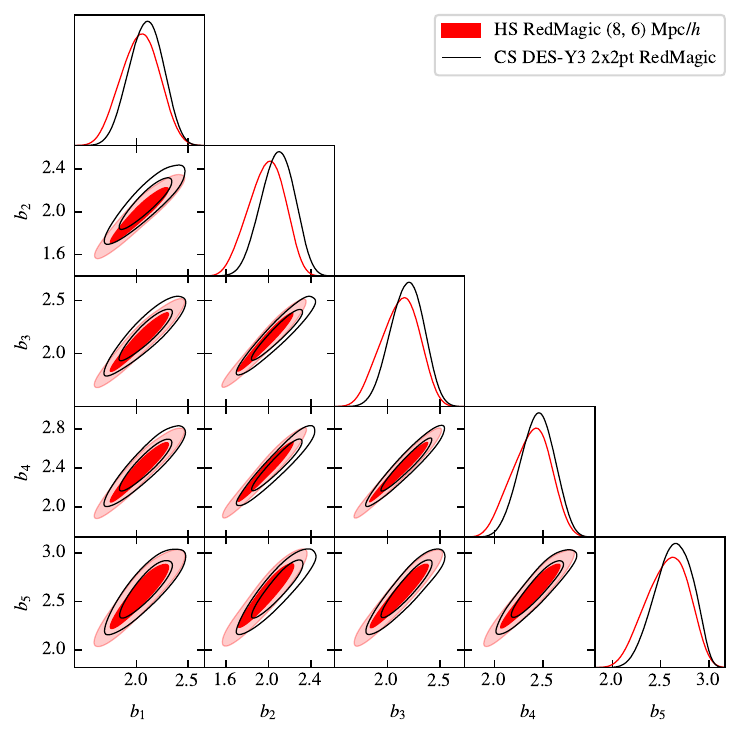}
  \caption{
  Marginalised posteriors of the linear galaxy bias found in the \redmagic analysis. The red, shaded contours are the HS results under the extended scale cut, and the black contours are the CS results from \protect\cite{2x2_redmagic}.
  }
  \label{fig:redmagic_bias}
\end{figure}

\begin{figure}
  \centering
  \includegraphics[width=0.99\linewidth]{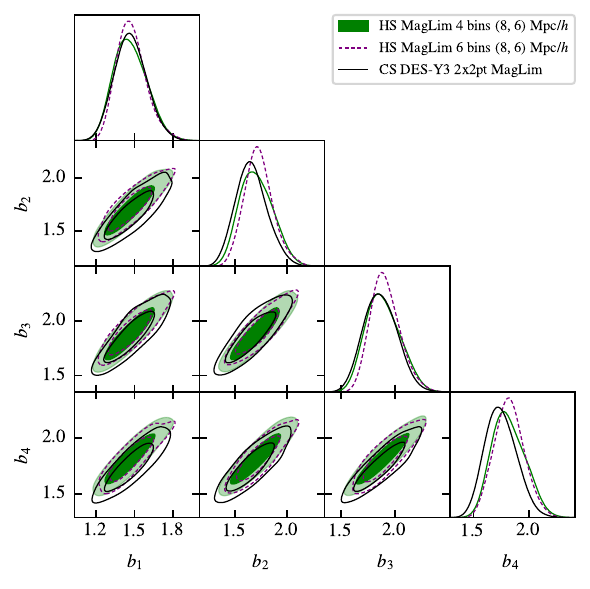}
  \caption{
  Marginalised posteriors of the linear galaxy bias found in the \maglim analysis. The green, shaded contours are the HS results under the extended scale cut and using the first four bins of the sample, while the purple, dashed curves are the constraints from using all six bins of \maglim. The black contours are the CS results from \protect\cite{2x2_maglim}.
  }
  \label{fig:maglim_bias}
\end{figure}

The marginalised posteriors of the galaxy bias under the extended scale cut are shown in Figures \ref{fig:redmagic_bias} and \ref{fig:maglim_bias}. We find an overall good compatibility between the harmonic and configuration space constraints. The \redmagic HS results systematically show a small tail towards lower values of $b_i$, making the constraints slightly looser than in CS. For \maglim, on the other hand, the results for the four bins analysis reproduce well what was found in CS, while the six bins run shows an expected increase in the constraining power. Appendix \ref{app:1d-post} shows the 1D marginalised posteriors of all parameters, along with their priors and their CS counterparts. A discussion about the goodness-of-fit of all these runs can be found in Appendix \ref{app:chi2}.

\subsection{wCDM}
\label{sec:wcdm}

\begin{figure}
  \centering
  \includegraphics[width=1.0 \columnwidth]{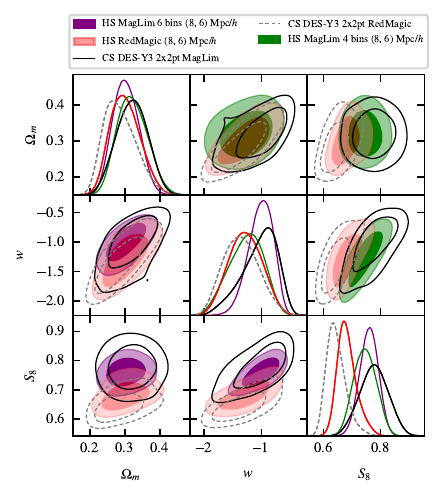}
  \caption{
    marginalised constraints on $w$, $S_8$ and $\Omega_m$ in $w$CDM from galaxy clustering and galaxy-galaxy lensing joint analyses. Purple and red shaded contours represent the harmonic space (HS) chains under the (8,6) Mpc/$h$ scale cut respectively. Solid and dashed contours represent the configuration space (CS) $w$CDM results from \protect\cite{2x2_redmagic} and \protect\cite{2x2_maglim}, which includes shear ratio measurements in their analysis and 4 tomographic bins of lenses for \maglim. }
  \label{fig:wcdm}
\end{figure}

\begin{figure*}
\begin{multicols}{2}
\includegraphics[width=0.94\linewidth]{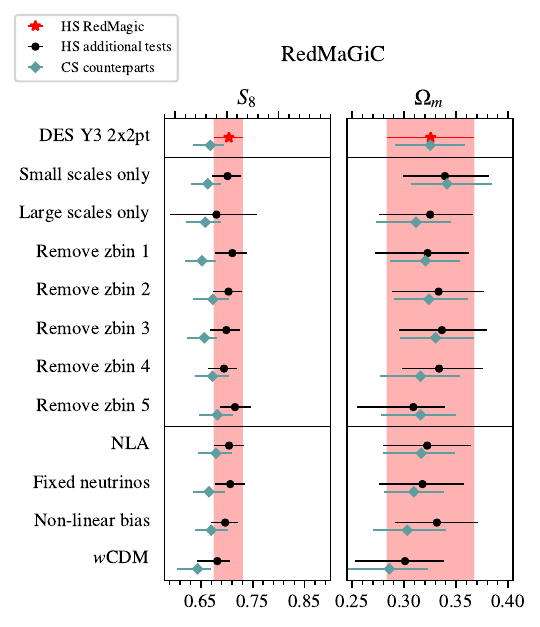}\par
\includegraphics[width=\linewidth]{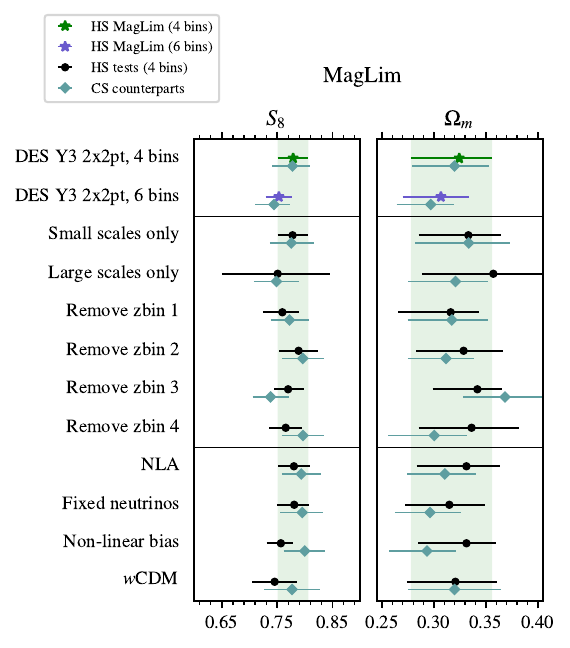}\par
\end{multicols}
\caption{
Internal consistency and robustness tests for the galaxy clustering and galaxy-galaxy lensing joint analyses for the extended scale cut. Left: \redmagic DES Y3 2$\times$2pt (red star) in harmonic space (HS), additional tests in HS (black circles) and their configuration space (CS) counterparts (green diamonds) from \protect\cite{2x2_redmagic}. Right: \maglim DES Y3 2$\times$2pt (purple star) in HS, additional tests in HS (black circles) and their CS counterparts (green diamonds) from \protect\cite{2x2_maglim}. All CS tests include shear ratio measurements and use 4 tomographic bins of lenses for \maglim, except where otherwise is mentioned. From top to bottom of both figures: first quadrant demarcates main results; second quadrant, internal consistency tests; and third quadrant, tests for the model robustness.
}
\label{fig:robustness}
\end{figure*}

Similarly to what was presented in Section \ref{sec:lcdm}, the harmonic space results of the analogous chains but considering the $w$CDM cosmological model and our extended scale cut are shown in Figure~\ref{fig:wcdm} (again, bottom triangle includes all six tomographic bins of \maglim, while the upper triangle includes only the first four bins). The HS constraints are represented by the shaded contours, and we also included their respective counterpart results in CS following the fiducial choices of \cite{2x2_redmagic} and \cite{2x2_maglim} (black and grey lines). 

The constraints derived for 
$\Omega_m$, $S_8$, $\sigma_8$, and $w$ can be seen in Table \ref{tab:results}, which also shows the agreement between harmonic and configuration space results in the $S_8 \times \Omega_m$ plane. All $w$CDM results in HS are compatible with CS under $1\sigma_{\text{2D}}$.
However, we note that the \redmagic and \maglim four bins constraints are very consistent with each other and are compatible with $w = -1$ within $1\sigma$.

\subsection{Internal consistency and robustness tests}
\label{sec:robustness_tests}

A series of additional tests on data was performed to investigate the internal consistency of the data vectors and the robustness of our modelling. The results of those tests are summarised in Figure~\ref{fig:robustness} for \redmagic (left) and \maglim (right). 

The first set of tests (second quadrant in Figure~\ref{fig:robustness}) consists in additional runs changing the structure of the data vectors. The \textit{small scales only} test restricts the analysis to physical scales smaller than the threshold of 30 Mpc/$h$ (same threshold used in \cite{2x2_redmagic} and \cite{2x2_maglim}), while \textit{large scales only} restricts the analysis to scales larger than this limit. The other consistency tests are chains removing the information from one tomographic bin of lenses at a time.

The second set of tests (third quadrant in Figure~\ref{fig:robustness}) corresponds to the following changes in the theoretical modelling:

\begin{itemize}
    \item NLA: variation of the intrinsic alignment model, setting parameters $a_2$, $\eta_2$ and $b_{\text{TA}}$ to zero.
    
    \item Fixed neutrinos: the mass range of neutrinos modelling is fixed at its lower boundary.
    
    \item Non-linear bias: the modelling also includes the non-linear galaxy bias effect, following the parameters in Table \ref{tab:parameters}.

    \item $w$CDM: changes the dark energy parametrization.
\end{itemize}

The respective results in configuration space for each internal consistency and robustness test are also shown in Figure~\ref{fig:robustness}, all following their main analyses choices of including shear ratios and four bins of lenses for \maglim (green diamond marker). 

Overall, these series of tests show that harmonic space analyses respond well to changes in the data vector and in the theoretical modelling. One can see that the large scales only results have much bigger error bars for $S_8$ than its configuration space counterpart. This is probably due to the fact that the $S_8$ information comes mainly from large multipoles $\ell$, which is also the reason why the extended scale cut gives tighter constraints for $S_8$ over the conservative, but does not improve $\Omega_m$ (see Figure~\ref{fig:chains}). Moreover, the limit of 30 Mpc$/h$ used to define small and large scales did not separate evenly the number of points in each group. Because of that, the large scales test had $1-4$ less points than the small scales test for each combination.


\section{Conclusions}
\label{sec:conc}

This work presents the harmonic space joint analysis of galaxy clustering and galaxy-galaxy lensing using the DES Y3 data.
We performed two independent analyses using the \redmagic and \maglim catalogues for lens galaxies. In both, the source galaxies came from the \metacalibration catalogue.

In order to compare these analyses with their counterparts in configuration space, the theoretical modelling and prior choice followed the ones described in \cite{2x2_redmagic} and \cite{2x2_maglim}, although the overall methodology was designed around the harmonic case. In particular, our main results have an optimised scale cut for harmonic space and include two versions of the \maglim analysis: using only the first four tomographic bins of lenses and using all six bins.

We describe our data and measurements in Section \ref{sec:data}.
The estimators used to measure the angular power spectra followed the PCL/MASTER method \citep{2002ApJ...567....2H}, implemented via the \namaster code \citep{Alonso_2019}.
The entire measurement and analysis pipelines were first rigorously tested using a large set of log-normal realisations. 
The same set of realisations and contaminated versions of the baseline data vector with non-linear galaxy bias and baryonic effects were used to select and validate our scale cuts and theoretical covariance. We presented two sets of scale cuts: the conservative ($R_{\text{min}}^{\text{gcl}} = 8 \text{ Mpc}/h$ and $R_{\text{min}}^{\text{ggl}} = 12 \text{ Mpc}/h$) and extended ($R_{\text{min}}^{\text{gcl}} = 8 \text{ Mpc}/h$ and $R_{\text{min}}^{\text{ggl}} = 12 \text{ Mpc}/h$) approaches.

For our fiducial \maglim analysis (first four tomographic bins of lenses and the extended scale cut) in the $\Lambda$CDM modelling, we found a clustering amplitude of $S_8 = 0.779\pm 0.028$ and a matter energy density of $\Omega_m = 0.324^{+0.032}_{-0.047}$, corresponding to a constraining power increase of $14\%$ in the $S_8 \times \Omega_m$ plane compared to the main results in \cite{2x2_maglim}. The same analysis but using all six tomographic bins resulted in $S_8 = 0.753\pm 0.024$ and $\Omega_m = 0.307^{+0.027}_{-0.037}$, representing $64\%$ increase in constraining power.
It is worth noting, however, that the six-bin analysis exhibited the poorest goodness-of-fit and the greatest impact on the cosmological constraints following the update of the covariance, as detailed in Appendices~\ref{app:cov_update} and~\ref{app:chi2}.
Their respective results for a \wCDM modelling was 
$S_8 = 0.745\pm 0.039$, $\Omega_m = 0.321^{+0.040}_{-0.047}$ and $w = -1.26^{+0.34}_{-0.27}$ for the four-bin analysis, and $S_8 = 0.759\pm 0.032$, $\Omega_m = 0.302\pm 0.036$ and $w = -1.01^{+0.24}_{-0.18}$ for the six-bin.

For \redmagic with the (8,6) Mpc/$h$ scale cut, on the other hand, our $\Lambda$CDM results were $S_8 = 0.704\pm 0.029$ and $\Omega_m = 0.325 \pm 0.040$, representing a $15\%$ decrease in the constraining power when compared to the main results in \cite{2x2_redmagic}. Interestingly, however, the harmonic chains for \redmagic prefer slightly higher values of $S_8$. 

The goodness-of-fit of all analyses are discussed in Appendix \ref{app:chi2} and the 1D marginalised posteriors of each parameter are shown along side with their priors and configuration space counterparts in Appendix \ref{app:1d-post}.

Additional tests were made and presented in Section ~\ref{sec:robustness_tests} to check the internal consistency of our data vectors and the robustness of the theoretical modelling. The tests had a very consistent agreement with our DES Y3 2$\times$2pt results
for \redmagic and \maglim, as Figure~\ref{fig:robustness} summarises.

This work is the first publication of the Dark Energy Survey collaboration to describe a methodology for galaxy-galaxy lensing in the harmonic space, paving the path for a $3\times 2$pt analysis in harmonic space for DES Y6, a project already under development.
To do so, other than combining the present work with the one developed for the DES Y3 cosmic shear in harmonic space \citep{Doux_2022} and incorporating the Y6 guidelines being currently developed for the 3$\times$2pt in configuration space, there are additional procedures to be worked on. 

We expect the methodology presented in this work to pave the way for and incentivise future harmonic space analyses of significantly larger datasets from next generation Stage-IV imaging programmes, such as those anticipated from LSST \cite{LSST2019} and Euclid \citep{EUCLID2011}.

\section*{Acknowledgements}

{\it Author Contributions:}
We acknowledge the invaluable contributions of everyone involved in this work.
LF contributed to the development and validation of the measurements, covariances and analysis pipelines, development of the manuscript, ran the pipeline, and coordinated the harmonic space team. 
FA-O contributed to the development of the measurements and analysis pipelines, and contributed to the supervision of LF and the development of the manuscript.
HC contributed to the development of the measurement and analysis pipeline, the development of the manuscript, and coordinated the harmonic space team.  
CD coordinated the team in early stages of this work, contributing to first versions of pipelines and the construction and validation of the \metacalibration catalogue.
RR provided helpful feedback that was instrumental in shaping the analysis and contributed to the supervision of LF.
ML contributed to the supervision of LF and offered valuable feedback which significantly informed the development of the analysis.
XF contributed with the original version of the \cosmolike code for computing the non-Gaussian contributions to the covariance matrix.
JP contributed to the construction and validation of the \metacalibration catalogue and its photometric redshifts and served as internal reviewer.
AP contributed to the construction and validation of the LSS catalogues and photometric redshifts and served as internal reviewer.
\textit{Construction and validation of the DES Y3 Gold catalog}:
KB, MB, MC-K, AC-R, AD-W, RG, ER, IS-N, ES, BYa.
\textit{Construction and validation of the DES Y3  \metacalibration catalogue}: 
AAm, MB, GB, CC, RC, AC, JE-P, AF, MG, DG, IH, MJ, NM, JMc, AN-A, SP, RR, CS, LS, ES, TS, MT, IT.
\textit{Construction and validation of the DES Y3  LSS catalogues and systematics}: MC, AC-R, JE-P, PF, MR-M, NW. 
\textit{Construction and validation of photometric redshifts}: AAl, AAm, GB, AC, RC, JC, MC, JD, SD, JE-P, SE, MG, GG, DG, IH, NM, JMc, JMy, MR, AR, ER, CS, MT, BYi.
The remaining authors have made contributions to this paper that include, but are not limited to, the construction of DECam and other aspects of collecting the data; data processing and calibration; developing broadly used methods, codes, and simulations; running the pipelines and validation tests; and promoting the science analysis. 

This work was partially supported by the Laborat\'orio Interinstitucional de e-Astronomia (LIneA), the Brazilian funding agencies CNPq and CAPES, the Instituto Nacional de Ci\^{e}ncia e Tecnologia (INCT) e-Universe (CNPq grant 465376/2014-2) and the S\~{a}o Paulo State Research Agency (FAPESP).
LF is supported by CNPq.
HC is supported by FAPESP (19/04881-8) and CNPq (151411/2022-0). FA-O would like to thank FAPESP grant  2016/01343-7 in August 2022 where part of this work was done.
ML is supported by FAPESP and CNPq. 
The authors acknowledge the use of computational resources from LIneA, the Center for Scientific Computing (NCC/GridUNESP) of the S\~{a}o Paulo State University (UNESP), and from the National Laboratory for Scientific Computing (LNCC/MCTI, Brazil), where the SDumont supercomputer ({\tt sdumont.lncc.br}) was used.
This research used resources of the National Energy Research Scientific Computing Center (NERSC), a Department of Energy Office of Science User Facility using NERSC award HEP-ERCAP-0027266.

Funding for the DES Projects has been provided by the U.S. Department of Energy, the U.S. National Science Foundation, the Ministry of Science and Education of Spain, 
the Science and Technology Facilities Council of the United Kingdom, the Higher Education Funding Council for England, the National Center for Supercomputing 
Applications at the University of Illinois at Urbana-Champaign, the Kavli Institute of Cosmological Physics at the University of Chicago, 
the Center for Cosmology and Astro-Particle Physics at the Ohio State University,
the Mitchell Institute for Fundamental Physics and Astronomy at Texas A\&M University, Financiadora de Estudos e Projetos, 
Funda{\c c}{\~a}o Carlos Chagas Filho de Amparo {\`a} Pesquisa do Estado do Rio de Janeiro, Conselho Nacional de Desenvolvimento Cient{\'i}fico e Tecnol{\'o}gico and 
the Minist{\'e}rio da Ci{\^e}ncia, Tecnologia e Inova{\c c}{\~a}o, the Deutsche Forschungsgemeinschaft and the Collaborating Institutions in the Dark Energy Survey. 

The Collaborating Institutions are Argonne National Laboratory, the University of California at Santa Cruz, the University of Cambridge, Centro de Investigaciones Energ{\'e}ticas, 
Medioambientales y Tecnol{\'o}gicas-Madrid, the University of Chicago, University College London, the DES-Brazil Consortium, the University of Edinburgh, 
the Eidgen{\"o}ssische Technische Hochschule (ETH) Z{\"u}rich, 
Fermi National Accelerator Laboratory, the University of Illinois at Urbana-Champaign, the Institut de Ci{\`e}ncies de l'Espai (IEEC/CSIC), 
the Institut de F{\'i}sica d'Altes Energies, Lawrence Berkeley National Laboratory, the Ludwig-Maximilians Universit{\"a}t M{\"u}nchen and the associated Excellence Cluster Universe, 
the University of Michigan, the National Optical Astronomy Observatory, the University of Nottingham, The Ohio State University, the University of Pennsylvania, the University of Portsmouth, 
SLAC National Accelerator Laboratory, Stanford University, the University of Sussex, Texas A\&M University, and the OzDES Membership Consortium.

The DES data management system is supported by the National Science Foundation under Grant Numbers AST-1138766 and AST-1536171.
The DES participants from Spanish institutions are partially supported by MINECO under grants AYA2015-71825, ESP2015-88861, FPA2015-68048, SEV-2012-0234, SEV-2016-0597, and MDM-2015-0509, 
some of which include ERDF funds from the European Union. IFAE is partially funded by the CERCA program of the Generalitat de Catalunya.
Research leading to these results has received funding from the European Research
Council under the European Union's Seventh Framework Program (FP7/2007-2013) including ERC grant agreements 240672, 291329, and 306478.
We  acknowledge support from the Australian Research Council Centre of Excellence for All-sky Astrophysics (CAASTRO), through project number CE110001020.

This manuscript has been authored by Fermi Research Alliance, LLC under Contract No. DE-AC02-07CH11359 with the U.S. Department of Energy, Office of Science, Office of High Energy Physics. The United States Government retains and the publisher, by accepting the article for publication, acknowledges that the United States Government retains a non-exclusive, paid-up, irrevocable, world-wide license to publish or reproduce the published form of this manuscript, or allow others to do so, for United States Government purposes.

Based in part on observations at Cerro Tololo Inter-American Observatory, 
National Optical Astronomy Observatory, which is operated by the Association of 
Universities for Research in Astronomy (AURA) under a cooperative agreement with the National 
Science Foundation.

The analysis presented in this work made use of the software tools {\sc SciPy}~\cite{2020SciPy-NMeth}, {\sc NumPy}~\cite{2020NumPy-Array},  {\sc Matplotlib}~\cite{Hunter:2007}, {\sc CAMB}~\cite{CAMB_Lewis20,Lewis:2002ah,Howlett:2012mh}, {\sc GetDist}~\cite{Lewis:2019xzd}, {\sc Multinest}~\cite{Feroz_2008,Feroz_2009,Feroz_2019},  {\sc Polychord}~\cite{Handley_2015}, {\sc CosmoSIS}~\cite{Zuntz_2015}, {\sc Cosmolike}~\cite{Krause_2017} and {\sc TreeCorr}~\cite{Jarvis_2004}



\bibliographystyle{mnras_2author}
\bibliography{references} 




\appendix

\section{Impact of covariance update}
\label{app:cov_update}

The choice of cosmological parameters employed to calculate the analytical covariance matrix within a Bayesian framework can introduce a bias into the inferred parameter constraints if these parameters deviate significantly from the recovered values (see \citealp{Friedrich_2021} and references therein).
To mitigate bias, the protocol adopted in this and other DES studies involves re-running the analysis with the covariance matrix updated to the best-fit values obtained from the previous iteration.

\begin{figure}
  \centering
  \includegraphics[width=0.99\linewidth]{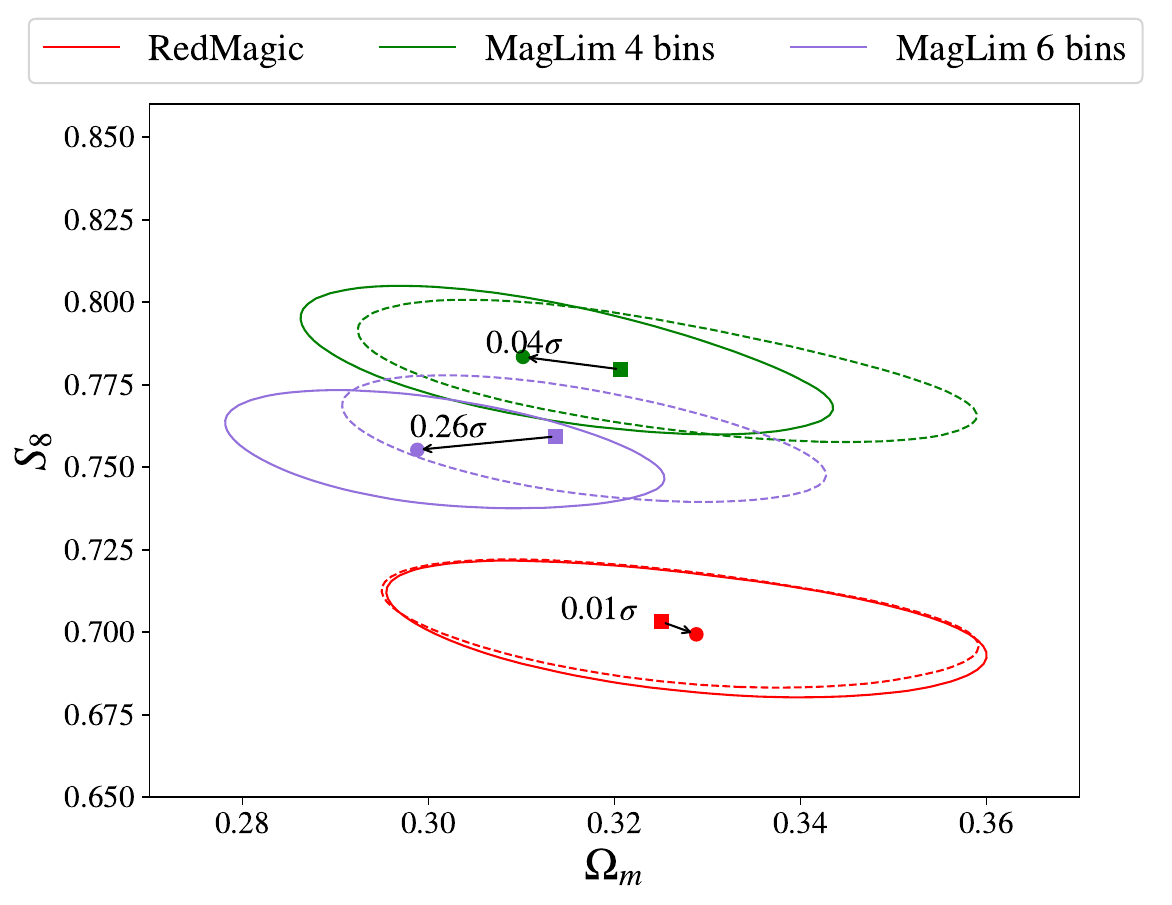}
  \caption{The impact on the $S_8 \times \Omega_m$ plane of updating the covariance for the different cases considered for the $2\times2$pt analysis in harmonic space. All cases assume the extended scale cut and \LCDM modelling.
  The square marker and dashed ellipses represent the best-fit value and $0.3\sigma_{\text{2D}}$ contour for the analyses using the original covariances (calculated at fiducial values of Table \ref{tab:parameters}).
  The circle marker and solid curves represent the analyses using the updated covariance (recalculated at the best-fit values of the previous run).
  }
  \label{fig:cov_update}
\end{figure}

Figure~\ref{fig:cov_update} illustrates the effect of covariance updates on the \LCDM analyses when applying the extended scale cut.
The second iteration of the covariance causes a low impact for \redmagic  ($0.01\sigma_{\text{2D}}$) and \maglim four bins ($0.04\sigma_{\text{2D}}$) in the $S_8 \times \Omega_m$ plane. For \maglim six bins, however, we see a greater impact, although still under $0.3\sigma_{\text{2D}}$. We tested updating one more time the covariance for \maglim six bins, but it did not result in a convergence as good as the others. As mentioned in \ref{sec:results}, the better convergence of the covariance and the goodness-of-fit (detailed in Appendix \ref{app:chi2}) were the reasons to choose the four bins configuration as the fiducial \maglim results of the present work.

\section{Measurements on the log-normal simulations}
\label{app:measurements}

\begin{figure*}
  \centering
  \includegraphics[width=1\textwidth]{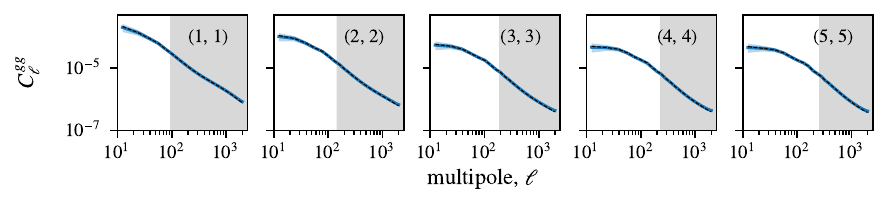}
  \centering
  \includegraphics[width=1\textwidth]{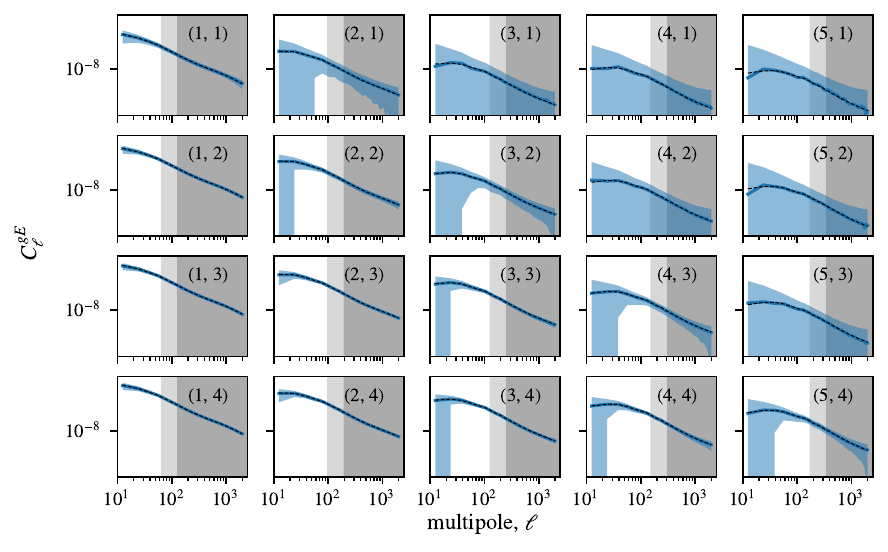}
  \caption{
  The angular power spectrum measurements for the set of 1800 log-normal realisations.
  Top panel: average of the galaxy clustering measurements $C_\ell^{\text{gg}}$ (solid line) and their standard deviation (blue, shaded area).
  The grey shaded area represents the scale cut based on the minimum physical scale $R^{\text{gcl}}_{\text{min}} = 8$ Mpc/$h$.
  Bottom panel: average of the galaxy-galaxy lensing (GGL) measurements $C_\ell^{\text{gE}}$ (solid line) and their standard deviation (blue, shaded area). The grey shaded area represents the scale cuts based on the minimum physical scale $R^{\text{ggl}}_{\text{min}} = 6$ Mpc/$h$ (light grey) and $R^{\text{ggl}}_{\text{min}} = 12$ Mpc/$h$ (dark grey).
  The indexes $(z_i, z_j)$ in each subplot indicate the redshift bin combination.
  For GGL, $z_i$ and $z_j$ refer to the bin of the lens and source, respectively.
  The black dashed curves show the input power spectra used to generate the log-normal realisations.
  }
  \label{fig:flask}
\end{figure*}

The pipeline developed to measure the angular power spectra from the DES catalogues was first validated using a set of 1800 log-normal, \redmagic-like realisations.
These realisations were generated using the \flask code \citep{flask}, as detailed in Section \ref{sec:Simslog-normal}. 

Figure \ref{fig:flask} presents the measurements of galaxy clustering (top panels) and galaxy-galaxy lensing (bottom panels).
It displays the mean of all realisations (solid, blue line) and the associated standard deviation (shaded, blue regions).
The simulated power spectra, assumed to be noiseless, used as input for generating the realisations are shown as the dashed black lines (see Table \ref{tab:parameters} for the fiducial parameters).
Across all tomographic combinations of power spectra, the simulated signal is faithfully reproduced by the mean of the realisations well within its standard deviation. 
This demonstrates the internal consistency of the measurement pipeline.

The same set of measurements on log-normal realisations was further used in the validation of the analytical covariance matrix and the analysis pipeline, as detailed in Section~\ref{sec:validation_with_simulation}.

\section{Goodness-of-fit}
\label{app:chi2}

In this section, we summarise and discuss the goodness-of-fit for all the \LCDM and \wCDM presented in Sections \ref{sec:lcdm} and \ref{sec:wcdm}.
As described in Section~\ref{sec:likelihood} and Appendix~\ref{app:cov_update}, the pipeline was run twice:
the first iteration used the original covariance matrix, calculated at the fiducial values listed in Table~\ref{tab:parameters}; and
the second, for each case, used an updated covariance matrix calculated at the corresponding first iteration best-fit values.
Table~\ref{tab:chi2} summarises the goodness-of-fit information for both iterations of each analysis.
It is important to consider that the presented reduced $\chi^2$ does not incorporate the informative priors.
Consequently, its actual value is likely to be lower, particularly for the simulations employing the conservative scale cut with fewer data points.
Appendix \ref{app:1d-post} presents the marginalised \LCDM posteriors alongside the priors for each parameter.

In particular, for the \maglim analyses employing all six bins of the sample, the update in the covariance matrix parameters leads to a significant increase in the $\chi^2$.
This effect, while less pronounced, mirrors the one already observed in the corresponding configuration space analysis, as described in Appendix B of \cite{2x2_maglim}.
Figure~\ref{fig:maglim_measurements} displays the residual plots of the \LCDM best-fit predictions against the angular power spectra measurements.

\begin{table*} 
  \caption{The goodness-of-fit for different configurations of the 2$\times$2pt analysis in harmonic space.
  The first column displays the cosmological model and the number of tomographic bins of lenses for the cases using the \maglim sample.
  The second column shows the scale cut used: either \textit{extended} (8, 6) Mpc/$h$ or \textit{conservative} (8, 12) Mpc/$h$.
  The covariance iteration column tells which version of the covariance was used: either original (calculated at the fiducial parameters shown in Table \ref{tab:parameters} or recalculated at best-fit values).
  The last five columns are: the number of total free parameters, the number of data points in the data vector after scale cuts, the $\chi^2$ at best-fit values, the reduced $\chi^2$ assuming the number of degrees of freedom (DoF) as number data points in the data vector minus the total number of parameters constrained, and the corresponding $p$-value.}
  \label{tab:chi2}
  \begin{tabular}{lccccccc} 
    \hline \hline
    \redmagic model & Scale cut & Cov. iteration & Total params. & data points & $\chi^2$ & $\frac{\chi^2}{\text{DoF}}$ & $p$-value  \\ 
    \hline \hline
    \LCDM & Extended & Updated & 30 & 227 & 187 & 0.95 & 0.68 \\
    \LCDM& Conservative & Updated & 30  & 159 & 128 & 0.99 & 0.51 \\
    $w$CDM& Extended & Updated & 31  & 227 & 190 & 0.96 & 0.63 \\
    \hline
    \LCDM&  Extended & Original & 30  & 227 &  206 & 1.05 & 0.32 \\
    \LCDM& Conservative & Original & 30  & 159 & 137 & 1.07  & 0.28 \\
    $w$CDM& Extended & Original & 31  & 227 &  210 & 1.09 & 0.18 \\
    
    \hline \hline
    \maglim model & Scale cut & Cov. iteration & Total params. & data points & $\chi^2$ & $\frac{\chi^2}{\text{DoF}}$ & $p$-value  \\ 
    \hline \hline
    \LCDM, 6 bins & Extended  & Updated & 37  & 295 & 314 & 1.22 & 0.01 \\
    \LCDM, 6 bins & Conservative & Updated & 37 & 207 & 230 & 1.35 & 0.001 \\
    $w$CDM, 6 bins & Extended & Updated & 38 & 295 & 304 & 1.17 & 0.03 \\
    \LCDM, 4 bins & Extended  & Updated & 31 & 178 & 146 & 0.99 & 0.51 \\
    \LCDM, 4 bins & Conservative  & Updated & 31 & 122 & 108 & 1.18 & 0.09 \\
    \wCDM, 4 bins & Extended  & Updated & 32 & 178 & 154 & 1.05 & 0.31 \\
    \hline
    \LCDM, 6 bins & Extended &  Original & 37 & 295 & 242 & 0.94 & 0.75 \\
    \LCDM, 6 bins & Conservative & Original & 37 & 207 & 178 & 1.05 & 0.32 \\
    $w$CDM, 6 bins & Extended & Original & 38 & 295 & 244 & 0.95 & 0.73 \\
    \LCDM, 4 bins & Extended &  Original & 31 & 178 & 135 & 0.92 & 0.77 \\
    \LCDM, 4 bins & Conservative  & Original & 31 & 122 & 94 & 1.03 & 0.40 \\
    \wCDM, 4 bins & Extended  & Original & 32 & 178 & 143 & 0.98 & 0.56 \\

    \hline
  \end{tabular}
\end{table*}

\section{Full marginalised posteriors}
\label{app:1d-post}

\begin{figure*}
  \centering
  \includegraphics[width=0.9\textwidth]{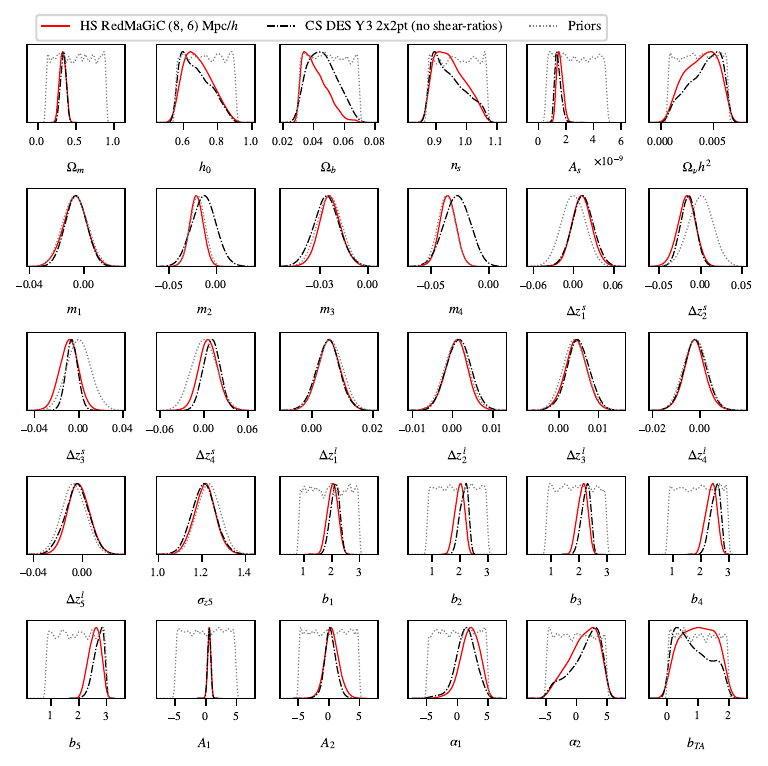}
    \caption{The full marginalised 1D posterior distributions for the fiducial harmonic space analysis of the \redmagic dataset (red).
    The fiducial DES Y3 configuration space analysis counterpart is also shown (black) for comparison as well as the input priors common to both analyses (grey).
    The prior distributions were sampled using the apriori sampler of \cosmosis.
    }
  \label{fig:full1d_posterior_redmagic}
\end{figure*}

\begin{figure*}
  \centering
  \includegraphics[width=0.9\textwidth]{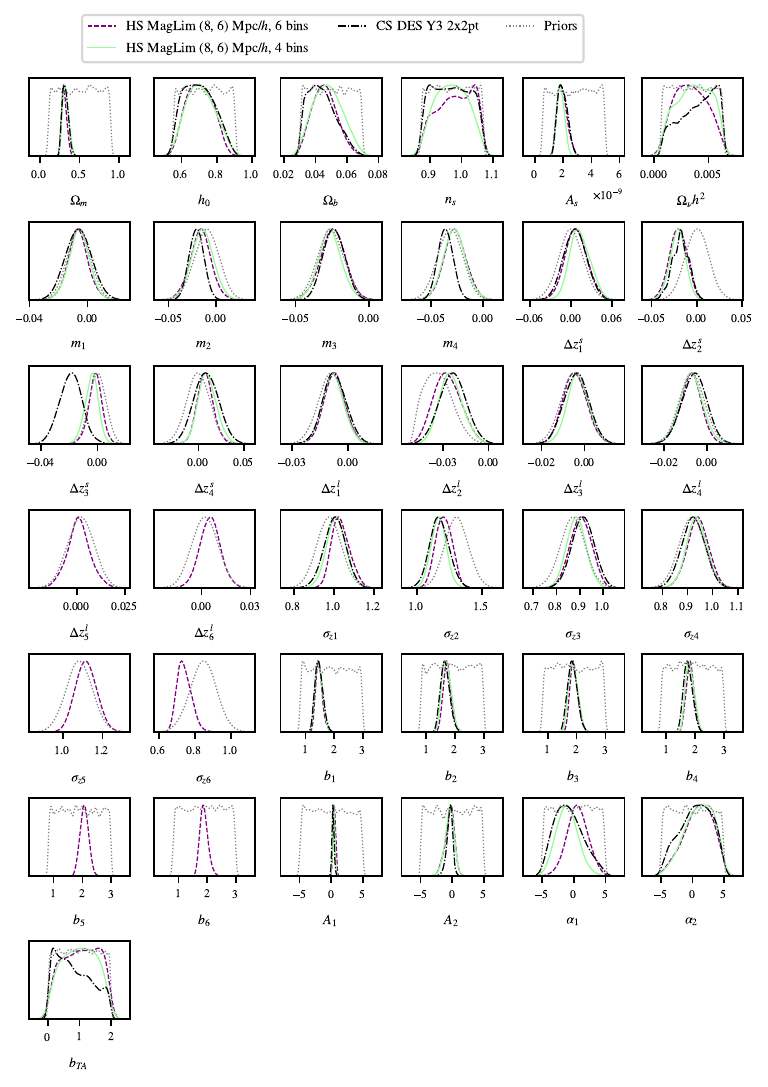}
  \caption{The full marginalised 1D posterior distributions for the fiducial harmonic space analysis of the \maglim dataset using 4 (purple) and 6 (green) tomographic bins of lenses. The DES Y3 configuration space analysis without shear ratios counterpart is also shown (black) for comparison as well as the input priors common to both analyses (grey).
  The prior distributions were sampled using the apriori sampler of \cosmosis.
    }
  \label{fig:full1d_posterior_maglim}
\end{figure*}

We show all the marginalised 1D posteriors for the full parameter space of our fiducial $\Lambda$CDM analysis in Figures~\ref{fig:full1d_posterior_redmagic} and \ref{fig:full1d_posterior_maglim} for \redmagic and \maglim, respectively.
For comparison, the configuration space counterparts from \cite{2x2_redmagic} and \cite{2x2_maglim} are also shown as well as their common prior distributions as computed by the apriori sampler\footnote{\url{https://cosmosis.readthedocs.io/en/latest/reference/samplers/apriori.html}} of \cosmosis.

We find no statistically significant constraints on the nuisance parameters, including photometric redshift biases and stretches, and cosmic shear calibration biases, beyond those imposed by the priors.

\section{Scale cuts in multipole space}
\label{app:scale_cuts}

In this appendix, we present the effective maximum multipoles, $\ell_{\rm max}$, associated with the different scale cut schemes discussed in Section \ref{sec:scale cuts}.
Table \ref{tab:scale_cuts} present summarises the different values for the different power spectra considered.

\begin{table*}
  \caption{Scale cuts used for the analyses. The maximum multipoles included were calculated from a certain minimum physical distance $R_{\text{min}}$, which is converted to $\ell_\text{max}$ via Equation \eqref{eq:lmax} for every photo-$z$ bin of lenses. This means that GGL combinations sharing the same lenses also share the same $\ell_{\text{max}}$. From this we derived our two sets of scale cuts. The conservative scale cut uses $R_{\text{min}} = (8, 12)$ Mpc$/h$ respectively for GCL and GGL, following the approach in \citep{desy1_3x2pt}. The extended scale cut, which is our fiducial choice, is derived from $R_{\text{min}} = (8, 6)$ Mpc$/h$.
  }
  \label{tab:scale_cuts}
  \begin{tabular}{lccc ccc} 
    \hline\hline
    & \multicolumn{3}{c}{\textbf{\redmagic + \metacalibration}} & \multicolumn{3}{c}{\textbf{\maglim + \metacalibration}} \\
    \cmidrule(lr){2-4}\cmidrule(lr){5-7}
    
      & GCL & GGL & GGL & GCL & GGL & GGL  \\ 
      & $R_{\text{min}} = 8$Mpc$/h$ & $R_{\text{min}} = 6$Mpc$/h$ & $R_{\text{min}} = 12$Mpc$/h$ & $R_{\text{min}} = 8$Mpc$/h$ & $R_{\text{min}} = 6$Mpc$/h$ & $R_{\text{min}} = 12$Mpc$/h$ \\
    \cmidrule(lr){2-4}\cmidrule(lr){5-7}
    
    Lens 1 & $\ell_{\text{max}} = 94 $ & $\ell_{\text{max}} = 126$ & $\ell_{\text{max}} = 63$  & $\ell_{\text{max}} = 105$ & $\ell_{\text{max}} = 139$ & $\ell_{\text{max}} = 70$\\
    Lens 2 & $\ell_{\text{max}} = 144$ & $\ell_{\text{max}} = 193$ & $\ell_{\text{max}} = 96$  & $\ell_{\text{max}} = 154$ & $\ell_{\text{max}} = 204$ & $\ell_{\text{max}} = 103$\\
    Lens 3 & $\ell_{\text{max}} = 187$ & $\ell_{\text{max}} = 250$ & $\ell_{\text{max}} = 125$ & $\ell_{\text{max}} = 199$ & $\ell_{\text{max}} = 264$ & $\ell_{\text{max}} = 133$\\
    Lens 4 & $\ell_{\text{max}} = 226$ & $\ell_{\text{max}} = 302$ & $\ell_{\text{max}} = 151$ & $\ell_{\text{max}} = 237$ & $\ell_{\text{max}} = 315$ & $\ell_{\text{max}} = 158$\\
    Lens 5 & $\ell_{\text{max}} = 255$ & $\ell_{\text{max}} = 340$ & $\ell_{\text{max}} = 170$ & $\ell_{\text{max}} = 265$ & $\ell_{\text{max}} = 353$ & $\ell_{\text{max}} = 177$\\
    Lens 6 & -                         & -                         & -                         & $\ell_{\text{max}} = 283$ & $\ell_{\text{max}} = 376$ & $\ell_{\text{max}} = 189$\\
    \hline
  \end{tabular}
\end{table*}

\section*{Affiliations}
$^{1}$ Departamento de F\'isica Matem\'atica, Instituto de F\'isica, Universidade de S\~ao Paulo, CP 66318, S\~ao Paulo, SP, 05314-970, Brazil\\
$^{2}$ Laborat\'orio Interinstitucional de e-Astronomia - LIneA, Rua Gal. Jos\'e Cristino 77, Rio de Janeiro, RJ - 20921-400, Brazil\\
$^{3}$ Department of Physics, University of Michigan, Ann Arbor, MI 48109, USA\\
$^{4}$ Brookhaven National Laboratory, Bldg 510, Upton, NY 11973, USA\\
$^{5}$ Instituto de F\'{i}sica Te\'orica, Universidade Estadual Paulista, S\~ao Paulo, Brazil\\
$^{6}$ ICTP South American Institute for Fundamental Research\\ Instituto de F\'{\i}sica Te\'orica, Universidade Estadual Paulista, S\~ao Paulo, Brazil\\
$^{7}$ Department of Physics and Astronomy, University of Pennsylvania, Philadelphia, PA 19104, USA\\
$^{8}$ Universit\'e Grenoble Alpes, CNRS, LPSC-IN2P3, 38000 Grenoble, France\\
$^{9}$ Department of Astronomy, University of California, Berkeley,  501 Campbell Hall, Berkeley, CA 94720, USA\\
$^{10}$ Department of Astronomy/Steward Observatory, University of Arizona, 933 North Cherry Avenue, Tucson, AZ 85721-0065, USA\\
$^{11}$ Department of Astronomy and Astrophysics, University of Chicago, Chicago, IL 60637, USA\\
$^{12}$ Nordita, KTH Royal Institute of Technology and Stockholm University, Hannes Alfv\'ens v\"ag 12, SE-10691 Stockholm, Sweden\\
$^{13}$ Ruhr University Bochum, Faculty of Physics and Astronomy, Astronomical Institute, German Centre for Cosmological Lensing, 44780 Bochum, Germany\\
$^{14}$ Argonne National Laboratory, 9700 South Cass Avenue, Lemont, IL 60439, USA\\
$^{15}$ Institute of Space Sciences (ICE, CSIC),  Campus UAB, Carrer de Can Magrans, s/n,  08193 Barcelona, Spain\\
$^{16}$ Fermi National Accelerator Laboratory, P. O. Box 500, Batavia, IL 60510, USA\\
$^{17}$ Department of Astrophysical Sciences, Princeton University, Peyton Hall, Princeton, NJ 08544, USA\\
$^{18}$ Institut de F\'{\i}sica d'Altes Energies (IFAE), The Barcelona Institute of Science and Technology, Campus UAB, 08193 Bellaterra (Barcelona) Spain\\
$^{19}$ Institute of Cosmology and Gravitation, University of Portsmouth, Portsmouth, PO1 3FX, UK\\
$^{20}$ Physics Department, 2320 Chamberlin Hall, University of Wisconsin-Madison, 1150 University Avenue Madison, WI  53706-1390\\
$^{21}$ University Observatory, Faculty of Physics, Ludwig-Maximilians-Universit\"at, Scheinerstr. 1, 81679 Munich, Germany\\
$^{22}$ Department of Physics \& Astronomy, University College London, Gower Street, London, WC1E 6BT, UK\\
$^{23}$ Department of Physics, Carnegie Mellon University, Pittsburgh, Pennsylvania 15312, USA\\
$^{24}$ NSF AI Planning Institute for Physics of the Future, Carnegie Mellon University, Pittsburgh, PA 15213, USA\\
$^{25}$ Instituto de Astrofisica de Canarias, E-38205 La Laguna, Tenerife, Spain\\
$^{26}$ Center for Astrophysical Surveys, National Center for Supercomputing Applications, 1205 West Clark St., Urbana, IL 61801, USA\\
$^{27}$ Department of Astronomy, University of Illinois at Urbana-Champaign, 1002 W. Green Street, Urbana, IL 61801, USA\\
$^{28}$ Institut d'Estudis Espacials de Catalunya (IEEC), 08034 Barcelona, Spain\\
$^{29}$ Physics Department, William Jewell College, Liberty, MO, 64068\\
$^{30}$ Kavli Institute for Cosmological Physics, University of Chicago, Chicago, IL 60637, USA\\
$^{31}$ Department of Physics, Duke University Durham, NC 27708, USA\\
$^{32}$ NASA Goddard Space Flight Center, 8800 Greenbelt Rd, Greenbelt, MD 20771, USA\\
$^{33}$ Jodrell Bank Center for Astrophysics, School of Physics and Astronomy, University of Manchester, Oxford Road, Manchester, M13 9PL, UK\\
$^{34}$ Hamburger Sternwarte, Universit\"{a}t Hamburg, Gojenbergsweg 112, 21029 Hamburg, Germany\\
$^{35}$ Lawrence Berkeley National Laboratory, 1 Cyclotron Road, Berkeley, CA 94720, USA\\
$^{36}$ Department of Physics and Astronomy, University of Waterloo, 200 University Ave W, Waterloo, ON N2L 3G1, Canada\\
$^{37}$ Jet Propulsion Laboratory, California Institute of Technology, 4800 Oak Grove Dr., Pasadena, CA 91109, USA\\
$^{38}$ Institute of Theoretical Astrophysics, University of Oslo. P.O. Box 1029 Blindern, NO-0315 Oslo, Norway\\
$^{39}$ SLAC National Accelerator Laboratory, Menlo Park, CA 94025, USA\\
$^{40}$ Instituto de Fisica Teorica UAM/CSIC, Universidad Autonoma de Madrid, 28049 Madrid, Spain\\
$^{41}$ School of Physics and Astronomy, Cardiff University, CF24 3AA, UK\\
$^{42}$ School of Mathematics and Physics, University of Queensland,  Brisbane, QLD 4072, Australia\\
$^{43}$ Santa Cruz Institute for Particle Physics, Santa Cruz, CA 95064, USA\\
$^{44}$ Center for Cosmology and Astro-Particle Physics, The Ohio State University, Columbus, OH 43210, USA\\
$^{45}$ Department of Physics, The Ohio State University, Columbus, OH 43210, USA\\
$^{46}$ Center for Astrophysics $\vert$ Harvard \& Smithsonian, 60 Garden Street, Cambridge, MA 02138, USA\\
$^{47}$ Australian Astronomical Optics, Macquarie University, North Ryde, NSW 2113, Australia\\
$^{48}$ Lowell Observatory, 1400 Mars Hill Rd, Flagstaff, AZ 86001, USA\\
$^{49}$ Centre for Gravitational Astrophysics, College of Science, The Australian National University, ACT 2601, Australia\\
$^{50}$ The Research School of Astronomy and Astrophysics, Australian National University, ACT 2601, Australia\\
$^{51}$ Department of Applied Mathematics and Theoretical Physics, University of Cambridge, Cambridge CB3 0WA, UK\\
$^{52}$ George P. and Cynthia Woods Mitchell Institute for Fundamental Physics and Astronomy, and Department of Physics and Astronomy, Texas A\&M University, College Station, TX 77843,  USA\\
$^{53}$ Kavli Institute for Particle Astrophysics \& Cosmology, P. O. Box 2450, Stanford University, Stanford, CA 94305, USA\\
$^{54}$ LPSC Grenoble - 53, Avenue des Martyrs 38026 Grenoble, France\\
$^{55}$ Instituci\'o Catalana de Recerca i Estudis Avan\c{c}ats, E-08010 Barcelona, Spain\\
$^{56}$ Instituto de F\'isica Gleb Wataghin, Universidade Estadual de Campinas, 13083-859, Campinas, SP, Brazil\\
$^{57}$ Observat\'orio Nacional, Rua Gal. Jos\'e Cristino 77, Rio de Janeiro, RJ - 20921-400, Brazil\\
$^{58}$ Department of Physics, University of Genova and INFN, Via Dodecaneso 33, 16146, Genova, Italy\\
$^{59}$ Department of Physics, Northeastern University, Boston, MA 02115, USA\\
$^{60}$ Centro de Investigaciones Energ\'eticas, Medioambientales y Tecnol\'ogicas (CIEMAT), Madrid, Spain\\
$^{61}$ Department of Physics and Astronomy, Stony Brook University, Stony Brook, NY 11794, USA\\
$^{62}$ School of Physics and Astronomy, University of Southampton,  Southampton, SO17 1BJ, UK\\
$^{63}$ Computer Science and Mathematics Division, Oak Ridge National Laboratory, Oak Ridge, TN 37831\\
$^{64}$ Institut de Recherche en Astrophysique et Plan\'etologie (IRAP), Universit\'e de Toulouse, CNRS, UPS, CNES, 14 Av. Edouard Belin, 31400 Toulouse, France\\


\bsp	
\label{lastpage}

\end{document}